\documentclass[longbibliography,reprint,onecolumn,12pt,amsmath,amssymbaps]{revtex4-1}
\usepackage{epsfig}
\usepackage{graphicx}
\usepackage{dcolumn}
\usepackage{bm}
\usepackage{color} 
\usepackage{ulem}
\newcommand{\ba}{\begin{array}}
\newcommand{\ea}{\end{array}}
\newcommand{\be}{\begin{equation}}
\newcommand{\ee}{\end{equation}}
\newcommand{\bea}{\begin{eqnarray}}
\newcommand{\eea}{\end{eqnarray}}
\newcommand{\bfig}{\begin{figure}}
\newcommand{\efig}{\end{figure}}
\newcommand{\Bl}{\Bigl}
\newcommand{\Br}{\Bigr}

\newcommand{\RE}{{\rm Re}\,}

\newcommand{\re}{{\rm e}}
\newcommand{\ri}{{\rm i}}

\newcommand{\dd}{{\rm d}}
\newcommand{\al}{\alpha}

\newcommand{\vep}{\varepsilon}
\newcommand{\vp}{\varphi}
\newcommand{\om}{\omega}

\newcommand{\din}{\displaystyle\int\limits}

\newcommand{\sech}{\mathop{\rm sech}\nolimits}
\begin{document}
\title{Hydrodynamic models of astrophysical wormholes. \\ The general concept}
\author{Semyon Churilov$^1$, Yury Stepanyants$^{2,3}$}
\affiliation{$^1$Institute of Solar-Terrestrial Physics of the
Siberian Branch of Russian Academy of Sciences, Irkutsk-33, PO Box
291, 664033, Russia, e-mail: {\color{blue} Churilov@iszf.irk.ru}; \\
$^2$School of Sciences,
University of Southern Queensland, QLD 4350, Australia;\\ $^3$Department of Applied Mathematics, Nizhny Novgorod
State Technical University, Nizhny Novgorod, 603950, Russia, e-mail: {\color{blue} Yury.Stepanyants@usq.edu.au}.}
%
%
\begin{abstract}
\hspace*{0.5cm}\\
We study hydrodynamic models of astrophysical wormholes when water waves can be amplified in the course of propagation between two critical points where wave and current speeds coincide. Such models can be realised in shallow laboratory ducts with variable width and depths. In this paper, we derive the basic set of equations for shallow water waves on the spatially variable flow in the duct of a variable cross-section and present the asymptotic analysis of solutions in the neighbourhood of the critical points. The critical points mimic either the black hole (BH) horizon if the flow transits from the subcritical to the supercritical regime, or the white hole (WH) horizon if the flow transits from the supercritical to the subcritical regime. We study then the wave propagation in the flow with two horizons when the flow transits first the BH horizon and then the WH one. The region between the horizons mimics a wormhole in general relativity. For the sake of completeness, we also study the successive transition through WH and BH horizons. The theoretical results are illustrated by numerical calculations of wave propagation in both these arrangements. It is shown that the wave amplification after passing the active zone between the horizons takes place in BH--WH arrangements only and can occur for different relationships between the subcritical and supercritical flow velocities. The frequency dependence of the amplification factor is obtained and quantified in terms of the velocity ratio within and outside the ``wormhole domain''. In the next paper, we plan to present exact solutions for the specific velocity profile.
\end{abstract}

\pacs{Valid PACS appear here}
\maketitle

\section{INTRODUCTION}
\label{sec:1}
Almost all scalar wave phenomena that exist in different media can be also reproduced or modelled by means of water waves. Moreover, the majority of wave effects can be observed in water by naked eyes. Since that time when Unruh \cite{Unruh81} demonstrated the similarity between the equations of general relativity describing wave phenomena in the proximity of black holes and acoustic waves in non-uniform flows, the study of such processes was developed into a whole scientific direction called the {\it analog gravity} (see, e.g. \cite{ArtBH, AnalogGrav, Unruh18} and references therein). One of the problems studied within this direction is modeling the motion of massless particles in the curved space-time containing event horizons, ergoregions and/or other regions with special properties inherent in general relativity. The analogy between the astrophysical phenomena and their counterparts in continuous media is based on the fact that for some types of waves (their list is permanently expanding), the equation describing wave propagation in a non-uniform medium has the form of the d'Alembert equation in the pseudo-Riemannian space with the metric tensor $G_{\mu\nu}$:
\begin{equation}
 \frac{1}{\sqrt{-G}}\,\frac{\partial}{\partial x^\mu}\left(\sqrt{-G}\,G^{\mu\nu}\,
 \frac{\partial\varphi}{\partial x^\nu}\right) = 0,
 \label{dAlam}
\end{equation}
where $G = \det G_{\mu\nu}$, \ \ $G^{\mu\lambda}G_{\lambda\nu} = \delta^\mu_\nu$, \ \
  $\delta^\mu_\nu$ is the unit tensor, and Einstein's rule of summation over the repeated indices is applied. This fact makes it possible to simulate various space-time structures by variation the conditions of wave propagation in laboratory experiments. 
  However, the aforementioned analogy is not too deep because the dynamical equations in hydrodynamics, which include the Navier--Stokes equation, and Einstein’s equations for space-time are not analogues. Nevertheless, many researchers employ the kinematic analogy to demonstrate that long water waves propagating on a flow can be treated as the gravitational analogue for the propagation of light in the curved space-time (see, for example, \cite{Peloquin, Euve16, Euve17, Euve20} and references therein). 

  In the pioneering paper by Unruh \cite{Unruh81}, for the modeling of a space-time structure, sound waves were proposed in the potential barotropic flow ${\bf U} = \nabla\Phi({\bf x}, t)$, where $\Phi$ is the hydrodynamic potential of the fluid velocity ${\bf U}$. Later it was recognised that instead of sound waves, surface gravity waves can be used on non-uniform flows in shallow water \cite{SchUn}. Despite the fact that, over the past three decades, the arsenal of analog gravity has expanded significantly, the study of the effects associated with the propagation of sound and water waves continues to receive much attention. This is confirmed, in particular, by the materials of the recently held discussion on the current state of research in the field of analog gravity organised under the auspices of the London Royal Society \cite{NextGen}. As was pointed out in the discussion, one of the topical problems in this field is the necessity of a further study of the effects related to the transitions between subcritical ($|{\bf U}| < c$) and supercritical ($|{\bf U}| > c$) flows, where $c$ is the wave speed in the medium (sound speed or speed of long linear waves on shallow water). The features of the wave dynamics in the regions with a different character of flows should be also studied in detail. This was one of the motivations of our work presented here.

 Recently we constructed analytical solutions \cite{ChErSt17} for the problem of surface wave transformation in a one-dimensional duct with a smoothly varying water velocity along the flow direction, $U(x) > 0$. All possible relationships between the water velocity $U(x)$ and wave velocity $c$ have been studied in that paper for linear waves in the shallow-water approximation. The transitions of the flow velocity from the subcritical to the supercritical and vice versa have been studied in detail on the basis of the solutions derived. The role of positive and negative energy waves in the supercritical flow has been also noted. However, the exact analytical solutions in Ref. \cite{ChErSt17} were based on two assumptions that impose significant restrictions on the problem formulation. One of the assumptions is related to the piece-wise linear velocity profile $U(x)$; this simplifies the derivation of analytical solutions but is hardly achievable in the laboratory set-ups. Another assumption presumes that the wave velocity $c$ is $x$-independent; such independence is difficult to achieve in the experiment too. In particular, in the experiments described in Refs. \cite{Euve15, Wein17, Euve20}, the water depth $H$ (and the surface wave velocity $c = \sqrt{gH\vphantom{^2}}$ (with $g$ being the acceleration due to gravity) was not maintained constant throughout the flow. Moreover, in some recent publications \cite{Aur15, Aur15a, Aur19}, for the simulation of the curved space-time, the new types of flows were proposed and realised in the laboratory. In those experiments, the current velocities $U$ remained constant, whereas the effective sound speed varied in space due to the special construction of the upper duct boundary.

 As follows from the aforementioned, to clarify the basic features of the wave-current interaction, there is a necessity to study analytically wave propagation in the case when both velocities, $U$ and $c$ are $x$-dependent in a rather arbitrary way. To this end, we can employ the method originally developed in the quantum theory of scattering (see, for example, \cite{Calo, Bab}). The method, in a certain sense, allows one to separate in each point of $x$ the contributions to the total disturbance of waves travelling in the opposite directions relative to the moving medium, This makes physically clearer both the calculations and results obtained. As in our previous paper \cite{ChErSt17}, we will consider surface water waves on the gradually varying flow in a shallow duct assuming that both the fluid depth $H(x)$ and the duct width $W(x)$ are slow functions of $x$.
 
\begin{figure}[b!]
\vspace*{6cm}%
\hspace*{-1.5cm}%
\begin{picture}(300,6)%
\multiput(76,18)(0,13){7}{\line(0,1){6}}
\multiput(255,18)(0,13){7}{\line(0,1){6}}
\put(60,50){{\Large $x_1$}}%
\put(258,50){{\Large $x_2$}}%
{\thicklines %
\put(-25,25){\large $U_1$}%
\put(-60,102){\vector(1,0){30}}%
\put(-60,82){\vector(1,0){30}}%
\put(-60,62){\vector(1,0){30}}%
\put(-60,42){\vector(1,0){30}}%
\put(-60,22){\vector(1,0){30}}%
\put(140,42){\large $U_2$}%
\put(105,82){\vector(1,0){30}}%
\put(105,72){\vector(1,0){30}}%
\put(105,62){\vector(1,0){30}}%
\put(105,52){\vector(1,0){30}}%
\put(105,42){\vector(1,0){30}}%
\put(335,25){\large $U_3$}%
\put(300,102){\vector(1,0){30}}%
\put(300,82){\vector(1,0){30}}%
\put(300,62){\vector(1,0){30}}%
\put(300,42){\vector(1,0){30}}%
\put(300,22){\vector(1,0){30}}%
{\color{blue}%
\put(350,52){\vector(1,0){30}}
\put(360,38){{\large $T$}}
\put(-10,82){\vector(1,0){30}}
\put(0,68){{\large $I$}}
\put(20,52){\vector(-1,0){30}}
\put(0,38){{\large $R$}}
\put(190,82){\vector(1,0){30}}
\put(200,68){{\large $P$}}
\put(190,52){\vector(1,0){30}}
\put(200,38){{\large $N$}}
}%
\put(-50,62){\vector(1,0){455}}
\put(396,46){{\Large $x$}}%
\put(28,-12){\vector(0,1){180}}
\put(33,160){{\Large $z$}}%
\put(31,65){\large $0$}%
\qbezier(26,122)(40,122)(56,114)
\qbezier(56,114)(96,89)(136,89)
\qbezier(190,89)(230,89)(270,115)
\qbezier(270,115)(280,120)(300,122)
\qbezier(26,2)(40,4)(56,10)
\qbezier(56,10)(96,30)(136,36)
\qbezier(190,36)(230,30)(270,9)
\qbezier(270,9)(280,2)(300,2)
\put(136,36){\line(1,0){54}}%
\put(136,89){\line(1,0){54}}%
\hspace*{-5mm}
\put(-50,122){\line(1,0){90}}%
\put(310,122){\line(1,0){90}}%
\put(-50,2){\line(1,0){90}}%
\put(310,2){\line(1,0){90}}%
}%
\hspace*{5mm}
\put(-6,127){{\large $W/2$}}%
\put(-17,-12){{\large $-W/2$}}%
{\color{red}
\put(0,105){\circle{20}}
\put(-3,102){\large $1$}%
\put(370,105){\circle{20}}
\put(367,102){\large $3$}%
\put(163,75){\circle{20}}
\put(160,72){\large $2$}%
}
\end{picture}
\vspace{5mm} %
\caption{(Color online) \protect\footnotesize Sketch of a duct of a variable width and depth with a non-uniform fluid flow. Positions $x_1$ and $x_2$ designate the coordinates of black and white hole horizons, respectively, where $U(x) = c(x)$. Region 2 mimics a model of a wormhole connecting black and white holes. For other notations see the text.}%
\label{f00}%
\end{figure}
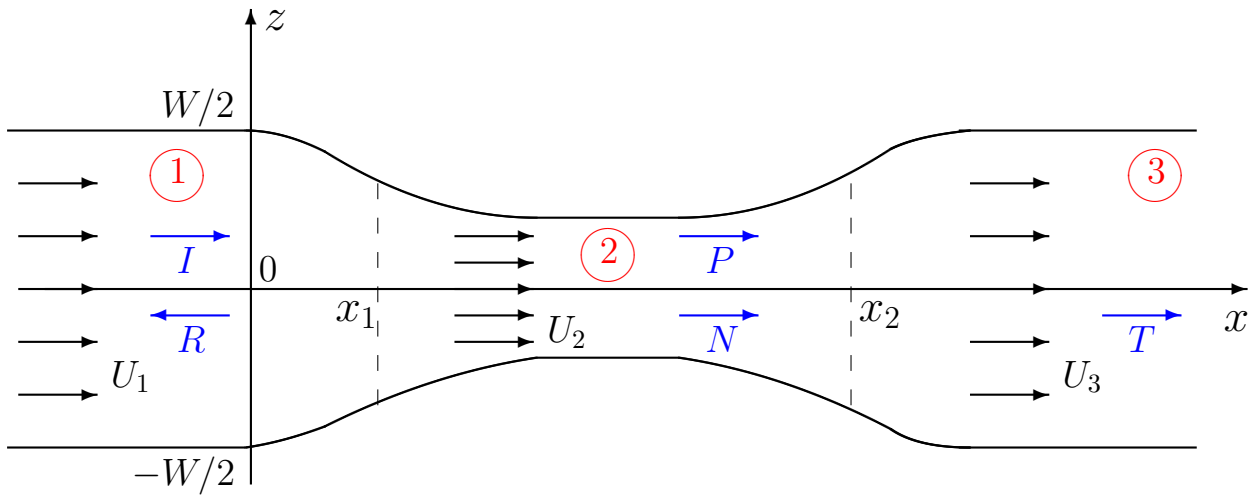

Being equipped with the developed powerful mathematical tool, we will tackle the problem of the laser effect in hydrodynamics which models the wave penetration through the wormhole in the theory of relativity. The term ``wormhole'' was introduced by John Wheeler in 1950th to describe a connection between the separate parts of the Universe where the black hole (the term promoted by J. Wheeler) is connected by the tunnel with its counterpart called {\it white hole} (the term introduce by Igor Novikov in 1960th). Despite the originally suggested wormholes were found unstable, it was shown later that their modification containing exotic matter possessing negative energy can stabilise wormholes \cite{Morris}. In such a case, a wormhole can be traversable in both directions. The wormhole analogue in hydrodynamics can stably exist in the portion of a duct where the fluid flow is supercritical as shown in Fig. \ref{f00}. 
There is no necessity in the existence of exotic matter, whereas negative energy waves can exist in this region \cite{FabrStep}. As the consequence of existence of stable analogues of wormholes, an analogue of black hole laser can be realised between the critical points of a current in a supercritical region of a flow. Note that the black hole laser effect was theoretically predicted by Corley and Jacobson \cite{Corley}, see also \cite{Leonhardt}. It was studied numerically in optics \cite{Faccio} and observed experimentally in the Bose--Einstein condensate \cite{Steinhauer}. 
In Ref. \cite{Peloquin} the authors studied numerically water wave packets on a spatially varying current in the presence of a surface tension. Using the {\it ad hoc} constructed equation describing water waves, they demonstrated that traversable and bidirectional analogue of wormholes do exist in fluid mechanics. However, surface waves in fluid with the surface tension are subject to dispersion which is absent in the general relativity. Therefore the analogy between water waves and electromagnetic waves becomes too far, although the physical effect of wave amplification remains qualitatively similar. In our paper, we consider the amplification of shallow water waves without a dispersion within the framework of real hydrodynamic equations. We assume that the flow velocity $U(x)$ can be controlled by a proper variation of the duct cross-section, both by the variation of width and depth. The water depth variation solves the problem of a zero-frequency mode (a stationary free surface deformation) mentioned in \cite{Peloquin}. Another problem mentioned in \cite{Peloquin} and related to the rapid damping of {\it small-scale capillary waves} is not topical in our case which pertains to {\it long gravity waves}. Therefore, the viscosity effects can be neglected almost everywhere in the flow except the critical points (BH or WH horizons) where $U(x) = c$. The effect of viscosity in such points plays an important role and is studied in detail in the Appendix. We show that the wave amplification is directly related  to the existence of negative energy waves between the analogue horizons. In this paper, we develop the general approach to the description of linear waves on inhomogeneous currents and demonstrate the importance of viscous effects near the critical points. The developed theoretical concept is illustrated by numerical calculations with some particular profiles of the mean flow $U(x)$ and the wave speed $c(x)$. In the next paper, we plan to present the analytical results on wave amplification for the exactly solvable flow model with the piece-wise linear velocity profile $U(x)$.

 The article is organized as follows. The problem formulation is given in Section \ref{sec:2} and the basic equations are also derived there. Section \ref{sec:3} is devoted to the asymptotic analysis of solutions in the neighbourhood of the critical points, where the flow and wave velocities coincide. Such a point is called the BH horizon (``the black hole horizon'') if the flow transits from the subcritical to the supercritical regime passing through the point. If the flow transits from the supercritical regime to the subcritical regime passing through the point, then it is called the WH horizon (``the white hole horizon''). In Sections \ref{sec:4} and \ref{sec:5} we study the wave propagation in the flow with two horizons, when the flow transits from BH to WH (BH--WH duct) or from WH to BH (WH--BH duct). Section \ref{sec:6} contains the results of numerical calculations of wave propagation in both these arrangements illustrating the theoretical analysis presented in Sections \ref{sec:4} and \ref{sec:5}. It is shown that the wave amplification can occur when an incident wave produces a higher amplitude transmitted wave passing through the ``wormhole'' between the BH and WH horizons. The amplification factor is estimated in terms of the velocity ratio within and outside the wormhole. Section \ref{sec:7} is devoted to the discussion of the results obtained. In Appendix, we present calculations demonstrating the influence of the water viscosity on the wave transformation in the vicinity of critical points (BH or WH horizons).

\section{GOVERNING EQUATIONS AND THE PROBLEM STATEMENT}
\label{sec:2}
In the physical model of the flow used in this study, the velocity of surface wave propagation relative to moving water is $c(x) = \sqrt{gH(x)}$. To ensure the independence of $c(x)$ and $U(x)$, we assume that the flow occurs in a duct of the width $W$ which smoothly varies with $x$ as required by the law of mass flux conservation,
\be
U(x)H(x)W(x) = {\rm const}.
\label{Flux}
\ee

In the hydrostatic approximation describing the propagation of long waves in shallow water \cite{LL}, the pressure can be represented as $p = p_0 + \rho g (\eta - y)$, where $p_0$ is the atmospheric pressure, $\rho = $ \, const is the water density, and  $\eta(x,y,t)$ is the deflection of the water surface from the equilibrium position. Present the horizontal fluid velocity as ${\bf V} = \{U(x)+u(x,y,t),\,v(x,y,t)\}$, where $|u(x,y,t)| \ll U(x)$ and substitute this into the linearized Euler equation:
 \begin{eqnarray}
 \dfrac{\partial u}{\partial t} &+& \dfrac{\partial(Uu)}{\partial x} = -g\dfrac{\partial\eta}{\partial x}, \label{Euler1} \\
 \dfrac{\partial v}{\partial t} &+& U\,\dfrac{\partial v}{\partial x} = -g\dfrac{\partial\eta}{\partial y}.
 \label{Euler2}
 \end{eqnarray}

Equation of mass conservation for shallow-water waves is:  \begin{equation}
\dfrac{\partial S}{\partial t}+\dfrac{\partial}{\partial x}\Bl[S(U+u)\Br]+
 \dfrac{\partial}{\partial y}\Bl(Sv\Br)=0,
 \label{Mass}
\end{equation}
where $S(x,y,t) = [H(x) + \eta(x,y,t)]W(x)$ is the part of the duct cross section occupied by water. Linearization of Eq. (\ref{Mass}) with respect to small perturbations $u$, $v$, and $\eta$ leads to the equation:
\begin{equation}
 \dfrac{\partial\eta}{\partial t}+HU\dfrac{\partial}{\partial x}\left(\dfrac{u}{U}+
 \dfrac{\eta}{H}\right)+H\,\dfrac{\partial v}{\partial y}=0.
 \label{eta}
 \end{equation}

Now let us introduce the potential $\varphi$ for the perturbed fluid velocity ${\bf v} = (\partial\varphi/\partial x,\,\partial\varphi/\partial y)$. Then, we find from Eqs. (\ref{Euler1}) and (\ref{Euler2}):
\begin{equation}
-g\eta = \dfrac{\partial\varphi}{\partial t} + U\dfrac{\partial\varphi}{\partial x}.
 \label{eta2}
 \end{equation}
Combining Eqs. (\ref{eta}) and (\ref{eta2}) and bearing in mind that  $c^2(x) = gH(x)$, we arrive to the single equation describing long surface wave propagation in terms of the velocity potential $\varphi$:
 \begin{equation}
 \left(\dfrac{\partial}{\partial t}+U\dfrac{\partial}{\partial x}-2U\dfrac{c'}{c}\right)
 \left(\dfrac{\partial\varphi}{\partial t}+U\dfrac{\partial\varphi}{\partial x}\right)=
 c^2U\dfrac{\partial}{\partial x}\left(\dfrac{1}{U}\dfrac{\partial\varphi}{\partial x}\right)+
 c^2\dfrac{\partial^2\varphi}{\partial y^2},
 \label{WEq0}
 \end{equation}
where prime stands for the derivative with respect to $x$. Alternatively this equation can be presented in the `covariant' form:
 \begin{equation}
 \dfrac{\partial}{\partial t} \left(-\dfrac{1}{c^2U}\,\dfrac{\partial\varphi}{\partial t} -
 \dfrac{1}{c^2}\,\dfrac{\partial\varphi}{\partial x}\right) +
 \dfrac{\partial}{\partial x} \left(-\dfrac{1}{c^2}\,\dfrac{\partial\varphi}{\partial t} +
 \dfrac{c^2 - U^2}{c^2U}\,\dfrac{\partial\varphi}{\partial x}\right)+\dfrac{\partial}{\partial y}
 \left(\dfrac{1}{U}\,\dfrac{\partial\varphi}{\partial y}\right) = 0.
\label{quasitens}
 \end{equation}
Comparing this equation with Eq. (\ref{dAlam}), we obtain for corresponding `metric tensor'  (cf., e.g., \cite{SchUn})
 \bea
 G_{\mu\nu} & = & \dfrac{G^2_2}{c^2(x)U^2(x)}\left(
 \ba{ccc}
 -\Bl[c^2(x)-U^2(x)\Br] & -U(x) & 0 \\
 -U(x)                  & 1     & 0 \\
 0                      & 0     & 1
 \ea
 \right), \nonumber \\
 G^{\mu\nu} & = & \left[\dfrac{U(x)}{G_2}\right]^2 \left(
 \ba{ccc}
 -1    & -U(x)         & 0 \\
 -U(x) & c^2(x)-U^2(x) & 0 \\
 0     & 0             & c^2(x)
 \ea
 \right),
 \label{G}
 \eea
where $G_2 > 0$ is a constant chosen for the reason of convenience.

For the elementary monochromatic wave which is independent of $y$ and has frequency $\omega$,  $\vp=\phi(x)\re^{-\ri\,\om\, t}$, Eq. (\ref{quasitens}) reduces to the ODE:
 \be
 (c^2-U^2)\dfrac{\dd^2\phi}{\dd x^2}+\left[2U^2\dfrac{c'}{c}-
 (c^2+U^2)\dfrac{U'}{U}+2\ri\om U\right]\dfrac{\dd\phi}{\dd x}+
 \left(\om^2-2\ri\,\om\, U\dfrac{c'}{c}\right)\phi=0.
 \label{WEq}
 \ee
Assuming that $\om$ is a big parameter, one can readily construct the JWKB solution to this equation:
\be
 \phi(x)=\sqrt{c(x)U(x)}\left[A_+\exp\left(\ri\,\om \int 
 \dfrac{\dd x}{c(x)+U(x)}\right)+A_-\exp\left(-\ri\,\om \int 
 \dfrac{\dd x}{c(x)-U(x)}\right)\right].
 \label{WKB}
 \ee
This solution describes a superposition of two waves propagating in the opposite directions -- co-current and counter-current. Note that the second term in the right-hand side is singular at the critical points, where $U(x) = c(x)$. From the physical point of view, the singularities are caused by the blocking of the wave propagating against the current, so that its phase velocity and wavelength tend to zero at the critical point. Therefore, the consideration of the crossing through the critical point requires special attention.

Following \cite{Bab}, we look for an exact solution to Eq. (\ref{WEq}) in the form similar to (\ref{WKB}):
\bea
 \phi(x) &=& A(x) \left[\exp\left(\ri\,\om \int \dfrac{\dd x}{c(x)+U(x)}\right)
 +R(x) \exp\left(-\ri\,\om \int \dfrac{\dd x}{c(x)-U(x)}\right)\right] \nonumber\\
{} & \equiv & A\Bl(e_+ + Re_-\Br), \label{Sol}
 \eea
where functions $A(x)$ and $R(x)$ can be interpreted respectively as the complex amplitude of a co-current propagating wave and the relative amplitude of the counter-current propagating wave (i.e., ``the local reflection coefficient''); $e_\pm$ denote the corresponding exponential functions.

The method of variation of constants provides the condition which relates the functions $A(x)$ and $R(x)$ (see, e.g., \cite{Bender}). According to this method, only the exponential functions should be differentiated when we calculate the derivative $\dd\phi/\dd x$:
\be
 \dfrac{\dd\phi}{\dd x}=\ri\om A(x)\left[\dfrac{e_+}{c+U}-
 \dfrac{R(x)e_-}{c-U}\right].
 \label{dd}
 \ee
Then, functions $A(x)$ and $R(x)$ must satisfy the equation:
 \be
 \dfrac{\dd A}{\dd x}\Bl(e_+ + Re_-\Br)+A\dfrac{\dd R}{\dd x}\,e_- = 0.
 \label{AR0}
 \ee

Calculating $\dd^2\phi/\dd x^2$ in Eq. (\ref{dd}) and substituting the result into Eq. (\ref{WEq}), we find:
\be
 \dfrac{\dd A}{\dd x}\Bl[(c-U)e_+ - (c+U)Re_-\Br]-A\dfrac{\dd R}{\dd x}
 (c+U)e_- - A\left(c'+c\,\dfrac{U'}{U}\right)\Bl(e_+ - Re_-\Br)=0.
 \label{AR1}
 \ee
Equations (\ref{AR0}) and (\ref{AR1}) completely describe functions $A(x)$ and $R(x)$. By resolving these equations with respect to the derivatives, we obtain:
\bea
  \label{A}
 \dfrac{\dd A}{\dd x} & = & a(x)\Bl(1-R(x)\re^{-\ri\,\Phi(x)}\Br)A(x),\\
   \label{R}
 \dfrac{\dd R}{\dd x} & = & -\,a(x)\left(\re^{\ri\,\Phi(x)}-
 R^2(x)\re^{-\ri\Phi(x)}\right),
 \eea
where
 \bea
 a(x)&=&\dfrac{1}{2}\left(\dfrac{c'}{c}+\dfrac{U'}{U}\right)\equiv
 \dfrac{1}{2}\,\dfrac{\dd\ln\Pi(x)}{\dd x}\,, \quad \Pi(x)=c(x)U(x),  \quad \re^{\ri\,\Phi(x)} = \dfrac{e_+(x)}{e_-(x)},
 \nonumber \\
 \Phi(x) &=& \om \int \al(x)\dd x, \quad
 \al(x)\,=\,\dfrac{1}{c(x)+U(x)}+\dfrac{1}{c(x)-U(x)} = \dfrac{2c(x)}{c^2(x) - U^2(x)}.
 \label{aFi}
 \eea

The system of equations (\ref{A}) and (\ref{R}) has a number of useful properties that make the analysis of wave propagation simpler and more intuitive. Firstly, it is clear that if in some flow region function $\Pi(x) = $ \, const, then both $A(x) = $\, const and $R(x) = $\, const, i.e. waves in this region {\it do not experience reflection} despite that $c$ and $U$ depend on $x$.

Secondly, the problem is reduced to the nonlinear first-order ODE -- the Riccati equation (\ref{R}); after finding a solution to this equation, the amplitude equation for $A(x)$ (\ref{A}) is immediately integrated:
\be
 A(x) = A_0\sqrt{\Pi(x)}\exp\Bl[- \int \dd x\,a(x)\,R(x)\re^{-\ri\Phi(x)}
 \Br].
 \label{A2}
 \ee

Thirdly, it is easy to see that in the flow regions that do not contain critical points (where $U(x) = c(x)$) Eqs. (\ref{A}) and (\ref{R}) have the first integral:
 \be
 {\cal E}=\dfrac{|A(x)|^2}{\Pi(x)}\,\Bl[1-|R(x)|^2\Br] = {\rm const}.
 \label{En}
 \ee
This equation can be treated as the conservation of wave action which is equivalent in our case to the conservation of the pseudo-energy \cite{McIntyre, FabrStep}. According to this law, the amplitude of the incident wave $|A(x)|$ can be presented through the transformation coefficient $|R(x)|^2$ and the ``geometric factor'' of the flow $\Pi(x)$.

  For the further consideration, it is convenient to introduce the normalised amplitude that does not depend on the geometric factor, $D(x) = A(x)/\sqrt{\Pi(x)}$, and the normalised coefficient $r(x) = R(x)\re^{-\ri\,\Phi(x)}$ (recall that $R(x)$ is a complex-valued function). These functions satisfy the following equations (see Eqs. (\ref{A})--(\ref{aFi})):
\bea
  \label{D}
 \dfrac{\dd D}{\dd x} & = & -\,a(x)\,r(x)\,D(x), \\
  \label{r}
 \dfrac{\dd r}{\dd x} & = & -\,a(x)\Bl[1-r^2(x)\Br]-\ri\,\om\,\al(x)\,r(x).
 \eea
Then, Eqs. (\ref{Sol}), (\ref{A2}) and (\ref{En}) take the form:
 \bea
  \label{Sol2}
 \phi(x) & = & A(x) \Bl[1+r(x)\Br]\exp\left(\ri\,\om \int 
 \dfrac{\dd x}{c(x)+U(x)}\right), \\
  \label{A3}
 A(x) & = & A_0\sqrt{\Pi(x)}\exp\left(- \int \dd x\,a(x)\,r(x)\right), \\
  \label{En2}
 {\cal E} & = & \dfrac{|A(x)|^2}{\Pi(x)}\,\Bl[1-|r(x)|^2\Br]\equiv|D(x)|^2
 \Bl[1-|r(x)|^2\Br] = {\rm const}.
 \eea
It should be noted that if at some point of flow $|r(x)| < 1$, then this
  inequality holds throughout the entire flow due to the conservation law (\ref{En2}) (the same is true for $|r(x)| > 1$).

In some cases it is convenient to present a solution for $r(x)$ through the module and argument $r(x)=|r(x)|\,\re^{\ri\,\theta(x)}$ (note that $|r(x)| = |R(x)|$, where $R(x)$ is a complex-valued function); then for these quantities we have the equations:
 \bea
 \label{mr}
 \dfrac{\dd|r|}{\dd x} & = & -\,a(x)\Bl[1-|r(x)|^2\Br]\cos\theta(x),
 \\
 \label{tet}
 \dfrac{\dd\theta}{\dd x} & = &\dfrac{a(x)}{|r(x)|}\Bl[1+|r(x)|^2\Br]\sin\theta(x)-
 \om\,\al(x).
 \eea

In the next Section \ref{sec:3} we will consider in detail the solution in the vicinity of a critical point and transition through the critical point.

\section{Solution in the Vicinity of a Critical Point}
\label{sec:3}
\subsection{The asymptotic expansion}
\label{ssec:3.1}
Let $x = x_0$ be the critical point such that $U(x_0) = c(x_0) \equiv c_0$. We consider the general case when the curves $c(x)$ and $U(x)$ intersect without tangency, i.e. when the quantity $\mu = U'(x_0) - c'(x_0)$ is nonzero and, moreover, not small,  $\mu = O(1)$. Let us introduce a parameter $0 < \vep \ll 1$, put $x - x_0 = \vep\xi$, and use the notation $f_0 = f(x_0)$ for any function $f(x)$. Expanding functions $c(x)$ and $U(x)$ in the Taylor series in the vicinity of $x = x_0$ and replacing $x$ with $\xi$ in Eq. (\ref{r}), we obtain:
\be
 \label{Eqb}
 \xi\dfrac{\dd r}{\dd\xi} = \dfrac{\ri\,\om}{\mu}\,r + \vep\xi\left[
 \dfrac{c'_0 + U'_0}{2c_0}\Bl(r^2 - 1\Br) - \dfrac{\ri\,\om}{2}\left(
 \dfrac{1}{c_0} - \dfrac{c''_0 - U''_0}{\mu^2}\right)r\right] + \vep^2\xi^2 S  + \ldots,
 \ee
 where
 $$
S = \dfrac{1}{2}\left[\left(\dfrac{c''_0 + U''_0}{c_0} - 
 \dfrac{{c'_0}^2 + {U'_0}^2}{c^2_0}\right)\Bl(r^2 - 1\Br) + 
 \dfrac{\ri\,\om}{2}\left(\dfrac{c'_0 + U'_0}{c^2_0} + \dfrac{2}{3}\,
 \dfrac{c'''_0 - U'''_0}{\mu^2}+\dfrac{(c''_0 - U''_0)^2}{\mu^3}
 \right)r\right].
 $$

Let us look for a solution to this equation in the form $r=r^{(0)}+\vep r^{(1)}+\vep^2r^{(2)}+\dots$ and introduce a notation $\beta = \om/\mu$. In the zero order on the parameter $\vep$, we obtain:
\be
 \xi\dfrac{\dd r^{(0)}}{\dd\xi} = \ri\,\beta\,r^{(0)}, \qquad
 r^{(0)} = \tilde{B}_0\xi^{\ri\,\beta},\ \ \tilde{B}_0={\rm const}.
 \label{0}
 \ee
In the first order on this parameter, $O(\vep)$, the equation is:
 \be
 \xi\dfrac{\dd r^{(1)}}{\dd\xi} = \ri\,\beta\,r^{(1)}+\xi\left[
 \dfrac{c'_0 + U'_0}{2c_0}\Bl(\tilde{B}^2_0\xi^{2\,\ri\,\beta} - 1\Br) - 
 \dfrac{\ri\,\om}{2}\left(\dfrac{1}{c_0} - \dfrac{c''_0 - U''_0}{\mu^2}
 \right)\tilde{B}_0\xi^{\ri\,\beta}\right].
 \label{1}
 \ee
Solution to this equation can be readily derived:
\be
 r^{(1)}=\xi\Bl(B_{10}+B_{11}\xi^{\ri\,\beta} + B_{12}\xi^{2\,\ri\,\beta}\Br),
 \label{r1}
 \ee
 where
 \be
 B_{10}=-\dfrac{c'_0 + U'_0}{2c_0(1-\ri\beta)}, \quad
 B_{11}=-\dfrac{\ri\,\om}{2}\left(\dfrac{1}{c_0} - \dfrac{c''_0 - U''_0}
 {\mu^2}\right)\tilde{B}_0, \quad
 B_{12}=\dfrac{c'_0 + U'_0}{2c_0(1+\ri\,\beta)}\,\tilde{B}^2_0
 \label{B1}
 \ee

In the second order on the parameter $\vep$, $O(\vep^2)$, the equation is:
 \be
 \ba{l}
 \xi\dfrac{\dd r^{(2)}}{\dd\xi} = \ri\,\beta\,r^{(2)}+\xi\left[
 \dfrac{c'_0 + U'_0}{c_0}r^{(0)} - \dfrac{\ri\,\om}{2}\left(
 \dfrac{1}{c_0} - \dfrac{c''_0 - U''_0}{\mu^2}\right)\right]r^{(1)}
 \\ \ \\ 
 +\,\dfrac{\xi^2}{2}\left[\left(\dfrac{c''_0 + U''_0}{c_0} - 
 \dfrac{{c'_0}^2 + {U'_0}^2}{c^2_0}\right)\Bl({r^{(0)}}^2 - 1\Br) + 
 \dfrac{\ri\om}{2}\left(\dfrac{c'_0 + U'_0}{c^2_0} + \dfrac{2}{3}\,
 \dfrac{c'''_0 - U'''_0}{\mu^2}+\dfrac{(c''_0 - U''_0)^2}{\mu^3}
 \right)r^{(0)}\right].
 \ea
 \label{2}
 \ee
Solution to this equation we present in the form:
 \be
 r^{(2)}=\xi^2\Bl(B_{20}+B_{21}\xi^{\ri\,\beta}+B_{22}\xi^{2\ri\,\beta}
 +B_{23}\xi^{3\ri\,\beta}\Br),
 \label{r2}
 \ee
where
 \bea
 B_{20} & = & -\,\dfrac{1}{2(2-\ri\,\beta)}\left[\dfrac{c''_0 + U''_0}{c_0}
 - \dfrac{{c'_0}^2 + {U'_0}^2}{c^2_0} + \ri\,\om\left(
 \dfrac{1}{c_0} - \dfrac{c''_0 - U''_0}{\mu^2}\right)B_{10}\right],
 \nonumber \\
 B_{21} & = & \left[\dfrac{c'_0 + U'_0}{2c_0}\,B_{10} +
 \dfrac{\ri\,\om}{8}\left(\dfrac{c'_0 + U'_0}{c^2_0} + \dfrac{2}{3}\,
 \dfrac{c'''_0 - U'''_0}{\mu^2}+\dfrac{(c''_0 - U''_0)^2}{\mu^3}\right)
 \right]\tilde{B}_0 - \dfrac{\ri\,\om}{4}\left(\dfrac{1}{c_0} - 
 \dfrac{c''_0 - U''_0}{\mu^2}\right)B_{11},  \nonumber \\
 B_{22} & = & \dfrac{1}{2(2+\ri\beta)}\left[\left(\dfrac{c''_0 + U''_0}
 {c_0} - \dfrac{{c'_0}^2 + {U'_0}^2}{c^2_0}\right)\tilde{B}^2_0 + 
 2\,\dfrac{c'_0 + U'_0}{c_0}\tilde{B}_0 B_{11} - \dfrac{\ri\,\om}{2}
 \left(\dfrac{1}{c_0} - \dfrac{c''_0 - U''_0}{\mu^2}\right)B_{12}\right],
 \nonumber \\
 B_{23} & = & \dfrac{c'_0 + U'_0}{2c_0(1+\ri\,\beta)}\,\tilde{B}^2_0B_{12}. \nonumber
 \eea

The solution obtained consists of two parts. One of them is the sum of a slowly varying terms that vanish at the critical point; this part of the solution is represented by the Taylor series:
\be
 r_{sl}(x)=B_{10}(x-x_0)+B_{20}(x-x_0)^2 + \ldots.
 \label{Slow}
 \ee
Another part of the solution contains the terms with the coefficients $\tilde{B}_0$ and $B_{nm}$ ($m \ge 1$); these terms are strongly oscillating when $x \to x_0$ (see equations (\ref{0}), (\ref{r1}), and (\ref{r2})). Note that $B_{nm} \sim \tilde{B}^m_0$ and stress that $B_{nm}$ depend on $B_{i0}$, but $B_{i0}$ does not depend on $B_{nm}$ with $m \ge 1$ (this property holds in the higher orders of expansion too). Therefore, the slowly varying part of the solution is entirely determined by the local behaviour of functions $c(x)$ and $U(x)$ in the vicinity of the critical point and {\it does not depend} on the fast oscillating contribution. The latter is determined by both the boundary conditions and behaviour of functions $c(x)$ and $U(x)$ to the right or to the left of the point $x_0$. Note, by the way, that $\tilde{B}_0$ and the entire quickly oscillating contribution can be different at left and right sides of $x_0$, therefore, the found solution is reasonable to represent in the form:
\be
 r(x) = r_{sl}(x) + B^{(\pm)}_0 |x-x_0|^{\ri\,\beta}\Bl[1+O(x-x_0)\Br],
 \label{rx0}
 \ee
where the signs plus and minus pertain to the regions where $x > x_0$ and $x < x_0$, respectively.

The integrals in the equations (\ref {Sol2}) and (\ref {A3}) converge for $x \to x_0 \pm 0$, and the asymptotic expansions for $A(x)$ and $\phi(x)$ can be writen as:
\bea
 \label{asA}
 A(x) &=& c_0A^{(\pm)}_0\left[1+\dfrac{c'_0\!+\!U'_0}{2c_0}\,(x-x_0)\left(1-
 \dfrac{B^{(\pm)}_0}{1+\ri\,\beta}\,|x-x_0|^{\ri\,\beta}\right)+O(|x-x_0|^2)
 \right],
 \\
 \label{asfi}
 \phi(x) &=& c_0A^{(\pm)}_0\Bl[1+B^{(\pm)}_0|x-x_0|^{\ri\,\beta}+O(|x-x_0|)
 \Br]\re^{\ri\,\psi}, \quad \psi=\om\!\int^{x_0}\!\dfrac{\dd x}{c+U}\,.
 \eea
As the next step, we need to investigate how these solutions can be extended for the transition through the critical point. This will be done in the next subsection.
\subsection{Transition through the critical point}
\label{ssec:3.2}
Equation (\ref{asfi}) contains only the principal expansion terms. The first term in the square brackets refers to the co-current propagating wave, and the second term (that is rapidly oscillating) refers to the counter-current propagating wave. As shown in \cite{ChErSt17}, for the correct matching solutions through the critical point, it is necessary to take into account a small viscosity of the medium. The details of calculation of the viscous problem are presented in the Appendix, and here we summarise the final result. (Note in brackets that in dispersive media there are other possibilities to match the solutions through the critical point. The dispersion invokes causality arguments similar to those which were considered by Hawking \cite{Hawking-1975} and Unruh \cite{Unruh-1976} for astrophysical black holes. This option requires, however, assuming that the flow was uniform in the distant past.)

If the current passes from the subcritical regime into supercritical, then
\be
 A^{(+)}_0 = A^{(-)}_0, \qquad  B^{(+)}_0 = B^{(-)}_0=0.
 \label{sub}
 \ee
In other words, co-current traveling waves ``do not feel'' that the point is critical. The absence of counter-current running waves is due to the fact that on both sides of the critical point they propagate out of this point and therefore, cannot reach it. Hence, for such a transition we have (see Eqs. (\ref{Slow}), (\ref{rx0}), (\ref{En2}), and (\ref{mr})):
\be
 r(x_0)=0 \quad \mbox{and} \quad {\cal E}(x_0+0)={\cal E}(x_0-0).
 \label{sub1}
 \ee
Therefore, we conclude that if $a(x_0)\ne 0$ in Eqs. (\ref{mr}) and (\ref{tet}), then functions $r(x)$ and $\cos\theta$ change their signs in the course of transition through the critical point, i.e. $\theta(x_0+0)=\theta(x_0-0)\pm\pi$.

On the contrary, in the course of the transition from the supercritical to the subcritical regime, the counter-current propagating waves run to the critical point and, approaching it with a decreasing wavelength, are completely absorbed in its viscous neighbourhood. Their relative amplitudes $B^{(+)}_0$ and $B^{(-)}_0$ depend on their propagation prehistory in the {\it different} flow regions and therefore, are not related with each other, whereas $A^{(+)}_0 = A^{(-)}_0$. The energy flux in the course of the transition through the critical point is not generally conserved,
 \[
 {\cal E}(x_0-0)=\dfrac{|A^{(-)}_0|^2}{\Pi(x_0)}\Bl(1-|B^{(-)}_0|^2\Br) \ne
 {\cal E}(x_0+0)=\dfrac{|A^{(+)}_0|^2}{\Pi(x_0)}\Bl(1-|B^{(+)}_0|^2\Br),
 \]
 it can either decrease or increase (see \cite{ChErSt17}).
%
\section{Wave propagation in the BH--WH duct model}
\label{sec:4}
\subsection{The flow model and preliminary analysis}
\label{ssec:4.1}
\begin{figure}[h!]
\vspace*{-4.0cm}
\hspace*{4.0cm}\centerline{\includegraphics[width=25cm]{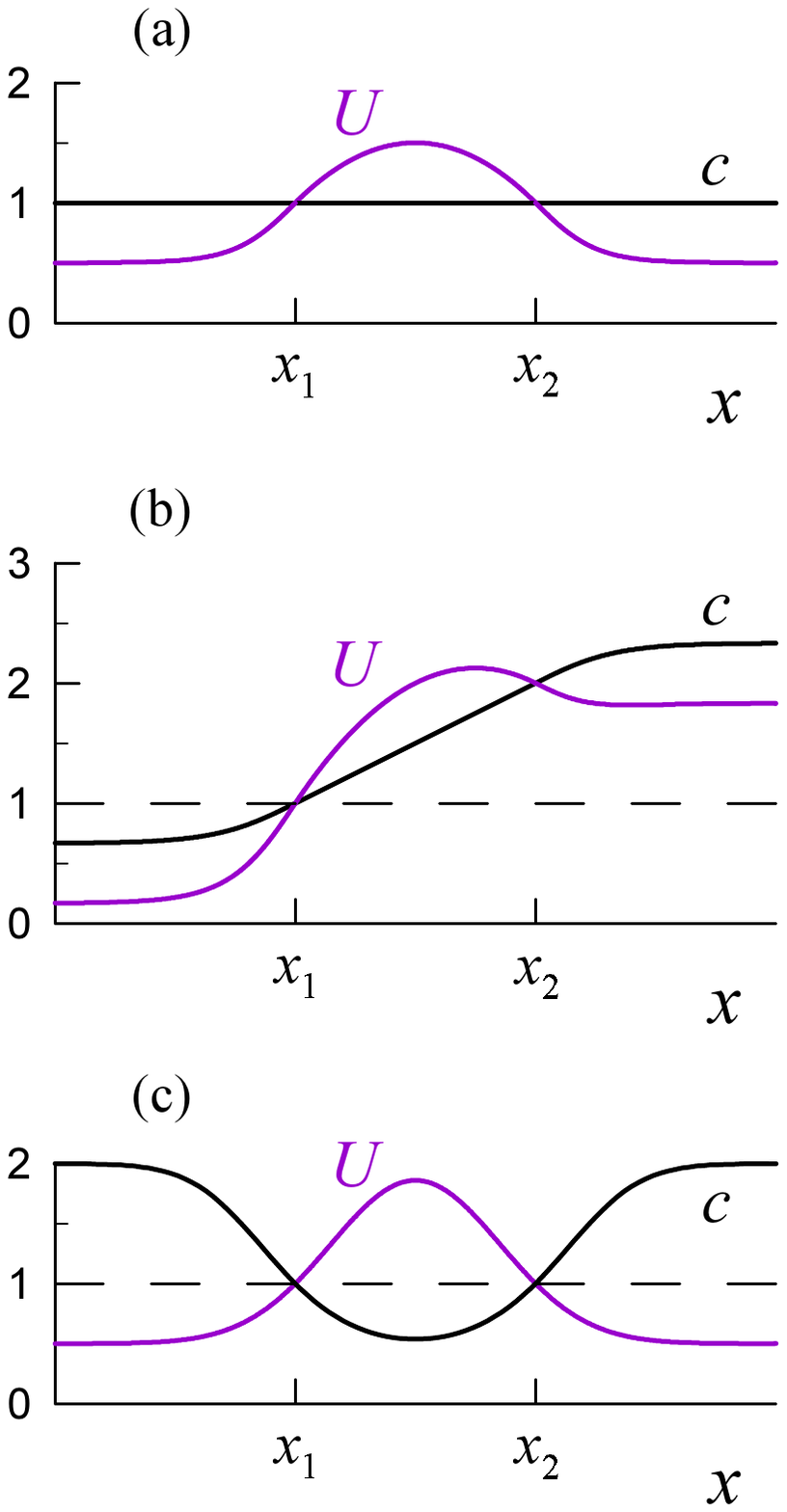}} \vspace*{-13.5cm}%
\caption{(Color online) \protect\footnotesize The flow models with two critical points: \ \ (a) -- flow (\ref{M1}), ${\cal M} = 1.5$; \ \ (b) -- flow (\ref{M2}), $c_2 = 2c_1 = 2$, $d = 1$; \ \ (c) --
reflectionless flow with $\Pi(x) \equiv c(x)U(x) = 1$.}
\label{f01}
\end{figure}

Consider the flow on the left ($-\infty < x < x_1$) and right ($x_2 < x < +\infty$) ends of which the flow is subcritical, $0 < U(x) < c(x)$, and within the middle part, for $x_1 < x < x_2$, -- supercritical, $U(x) > c(x)$ (see Fig. \ref{f01}). Here $x_1$ and $x_2$ are the critical points such that $U(x_1) = c(x_1) \equiv c_1$ and $U(x_2) = c(x_2) \equiv c_2$. In what follows, we will use the dimensionless variables such that:
\be
 c_1=1, \qquad \mu_1=U'_1-c'_1=1.
 \label{scale}
 \ee

Let us assume that a plane wave of frequency $\om$ arrives from the left. Its propagation in the inhomogeneous zone will be described in terms of functions $D(x)$ and $r(x)$. At the critical point $x = x_1$ and when $x \to +\infty$ there is no wave traveling upstream, therefore solution to Eq. (\ref{r}) must satisfy the boundary conditions:
 \be
 r(x_1)=0, \qquad r(+\infty)=0.
 \label{BC}
 \ee
Therefore, we can conclude that, firstly, in the entire flow $|r(x)| < 1$ (as it follows from the conservation law (\ref{En2})) and, secondly, the equation (\ref{r}) should be integrated into the upstream direction in the sub-critical areas and downstream -- on the supercritical section $x_1 \le x < x_2$. 

Further, suppose that for $x \to \pm \infty$ functions $c(x)$ and $U(x)$ tend to their nonzero limiting values, and let $D(-\infty) = 1$. Then from Eq. (\ref{A3}) and the matching conditions (\ref{sub}), $A^{(+)}_0 = A^{(-)}_0$, we find the amplitude of the downstream propagating wave as the function of $x$:
 \be
  \label{D1}
 D(x) = \exp\left[-\!\din_{-\infty}^{x}\!a(x')\,r(x')\,\dd x'\,\right], \quad
 -\infty<x<\infty.
 \ee
Then we can define the transmission ratio  $K(x)\!\equiv\!|D(x)|$ and present it as:
 \be
 \label{K}
 K(x) = \exp\left[-\!\din_{-\infty}^{x}\! 
 a(x')\RE\Bl(r(x')\Br)\,\dd x'\,\right] = \exp\left[-\!\din_{-\infty}^{x}\! a(x')|r(x')|\cos\theta(x')\,\dd x'\,\right].
 \ee
As can be seen from Eqs. (\ref{mr}) and (\ref{K}), the reflection coefficient $|r(x)|$ and transmission ratio $K(x)$ both increase or decrease simultaneously depending on the sign of the product $a(x)\cos{\theta(x)}$. Finally, in those regions where $a(x) = 0$, i.e. $\Pi(x) = $ const, they are constant too and, as already noted, waves running in the opposite directions do not interact, but the phase difference between them $\theta(x)$ changes due to the difference in the directions of propagation and difference in the wavelengths (i.e., because $\al(x) \ne 0$ -- see Eq. (\ref{tet})).

In accordance with the flow structure and conservation law (\ref{En2}), it is convenient to distinguish three stages for the downstream propagating wave. In the interval $x < x_1$, the transmission ratio $K(x)$ decreases from unity to $K(x_1) = \sqrt{1 - |r(-\infty)|^2}$. In the supercritical interval, the amplitude increases, and $K(x_2) = K(x_1)/\sqrt{1 - |r(x_2-0)|^2}$, and in the last interval, $x > x_2$, $K(x)$ decreases again to $K(+\infty) = K(x_2)\sqrt{1 - |r(x_2 + 0)|^2}$ (recall that in the general case $|r(x_2-0)| \ne |r(x_2 + 0)|$, therefore the increase and subsequent decrease of $K$ are independent). Thus, we can conclude that $K(x)$ can vary non-monotonically in any of these three intervals. As the result, in each interval of the flow, its integral transmission ratios are presented by the formulae:
$$
 K_1 = \sqrt{1-|r(-\infty)|^2} \le 1, \quad K_2 = \Bl[1-|r(x_2-0)|^2\Br]^{-1/2} \ge 1,
 $$
\be
 \label{K123}
K_3 = \Bl[1-|r(x_2+0)|^2\Br]^{1/2} \le 1,
 \ee
and the total transmission ratio (the transmission coefficient) is $T \equiv K(+\infty) = K_1 \cdot K_2 \cdot K_3$. It can be increased by (i) reducing the transformation of the co-current propagating wave into reflected waves in the sub-critical intervals of the flow and (ii) increasing  its transformation into the negative energy wave in the supercritical interval. Details of the concept of negative energy waves can be found in \cite{FabrStep, Maissa}. Here we only recall that a wave whose phase velocity is less than the local flow velocity is attributed as the negative energy wave. Such a wave cannot propagate against the flow because the flow is so strong that it pulls the wave in the direction of the flow.

The first condition can be fulfilled at once for all frequencies if in the subcritical regions the reflectionless wave propagation occurs with $a(x) \equiv 0$ (i.e. $c(x)U(x) \equiv $ const), then $K_1 = K_3 = 1$. Note that this condition does not prevent the change of the Mach number ${\cal M}(x) \equiv U(x)/c(x)$ along the flow, (see, e.g., Fig. \ref{f01}(c)). Another way to do it is based on the effect similar to that of anti-reflective coating in the optics. Its principle is that waves reflected from the different flow intervals add up in anti-phase and extinguish each other. As can be seen from Eq. (\ref{mr}), the quenching is complete if:
\begin{eqnarray}
 r(\!-\infty) &=& \!\din_{-\infty}^{x_1}\!a(x)\Bl[1\!-\!|r(x)|^2\Br]\cos\theta(x)\,\dd x\, = 0, \label{RF1} \\
 r(x_2\!+\!0) &=& \!\din^{\infty}_{x_2}\! a(x)\Bl[1\!-\!|r(x)|^2\Br]
 \cos\theta(x)\,\dd x\, = 0. \label{RF2}
 \end{eqnarray}
Since in those intervals where $a(x) = 0$, function $\theta(x)$ continues to vary, the fulfillment of Eqs. (\ref{RF1}) and (\ref{RF2}) can be ensured by variation of $\theta(x)$ due to ``inserts'' with $a(x) = 0$ of the required length in the proper intervals of the flow. But since for $a = 0$ \ \ $\dd \theta/\dd x \sim \om$ (see Eq. (\ref{tet})), the choice of the positions of such inserts and especially their lengths significantly depends on the wave frequency.

As will be shown below, both the selection of inserts and the difference in the speeds of wave propagation at the ends of the supercritical interval (both with $c_1 < c_2$ and $c_1 > c_2$) contribute to this, but the increase in wave frequency prevents to this.

To simplify further analysis, let us strengthen the previously formulated condition of fast convergence of $c(x)$ and $U(x)$ to their limiting values when $x \to \pm \infty$, assuming that $c$ and $U$ are constants and, respectively, $a(x) = 0$ for $x < x_- < x_1$ and $x > x_+ > x_2$. Below we perform the analysis for two limiting cases, when $\om\ll 1$ and $\om\gg 1$.
\subsection{The low-frequency limit, $\om\ll 1$}
\label{ssec:4.2}
Setting $\om = 0$ in Eq. (\ref{r}), we get:
\be
 \dfrac{\dd r_0}{\dd x}=-\dfrac{\Pi'(x)}{2\Pi(x)}\,\Bl(1-r^2_0\Br)\ \
 \Longrightarrow\ \ r_0(x)=\dfrac{R_0-\Pi(x)}{R_0+\Pi(x)}, \quad
 R_0=c_iU_i\,\dfrac{1+r_{0i}}{1-r_{0i}},
 \label{r0}
 \ee
where $c_i$, $U_i$, and $r_{0i}$ are the values of $c(x)$, $U(x)$, and $r_0(x)$ in some starting point $x = x_i$. If $x_i$ coincides with the first critical point $x_1$, then $r_{0i} = 0$, $R_0 = 1$ and
\be
 r_0(x)=\dfrac{1-\Pi(x)}{1+\Pi(x)}\equiv\dfrac{1-c(x)U(x)}{1+c(x)U(x)}=
 1-\dfrac{2c(x)U(x)}{1+c(x)U(x)}.
 \label{r01}
 \ee
Next, we look for a solution in the form: $r(x)=r_0(x)+\om\, r_1(x)+\om^2r_2(x)+\dots$. In the lowest order on the frequency, $O(\om)$, we have:
 \[
 \dfrac{\dd r_1}{\dd x}=\dfrac{\Pi'(x)}{\Pi(x)}\,r_0(x)r_1(x)-
 \ri\,\al(x)r_0(x).
 \]
Integrating this equation with the initial condition $r_1(x_i) = 0$ and using Eq. (\ref{r0}), we get:
 \be
 r_1(x)=-\dfrac{2\,\ri\,\Pi(x)}{[R_0+\Pi(x)]^2}\din_{x_i}^{x}\!
 \dfrac{r_0(\xi)}{U(\xi)}\,\dfrac{[R_0+\Pi(\xi)]^2}{c^2(\xi)-U^2(\xi)}\,\dd\xi\,.
 \label{r11}
 \ee
In the next order on the frequency, $O(\om^2)$, we have the following equation:
 \[
 \dfrac{\dd r_2}{\dd x}=\dfrac{\Pi'(x)}{2\Pi(x)}\Bl[2r_0(x)r_2(x)+
 r^2_1(x)\Br]-\ri\,\al(x)r_1(x).
 \]
Its solution subject to the initial condition $r_2(x_i)=0$ is:
 \be
 r_2(x)=-\dfrac{4\Pi(x)}{[R_0+\Pi(x)]^3}\din_{x_i}^{x}\!
 \dfrac{1}{U(\xi)}\,\dfrac{R_0\Pi(x)+\Pi^2(\xi)}{c^2(\xi)-U^2(\xi)}\,\dd\xi\,
 \din_{x_i}^{\xi}\!\dfrac{r_0(\eta)}{U(\eta)}\,
 \dfrac{[R_0+\Pi(\eta)]^2}{c^2(\eta)-U^2(\eta)}\,\dd\eta\,.
 \label{r21}
 \ee

The integral in the right-hand side of Eq. (\ref{r11}) can be evaluated in the vicinity of the point $x_2$; when $x\to x_2$ we have:
\[
 I(x)=\din_{x_i}^{x}\!\dfrac{r_0(\xi)}{U(\xi)}\,
 \dfrac{[R_0+\Pi(\xi)]^2}{c^2(\xi)-U^2(\xi)}\,\dd\xi = I_{1i} - \dfrac{R^2_0-c^4_2}{2\mu_2c^2_2}\,\ln{|x-x_2|} + O(|x-x_2|),
 \]
where $I_{1i} = $\;const, $\mu_2 = U'_2-c'_2$. The integral converges if $r_0(x_2) = 0$ (i.e. if $R_0 = c^2_2$), otherwise it logarithmically diverges.
 
 Similarly one can evaluate the integral in the right-hand side of Eq. (\ref{r21}):
 \[
 \ba{l}
 \din_{x_i}^{x}\!
 \dfrac{1}{U(\xi)}\,\dfrac{R_0\Pi(x)+\Pi^2(\xi)}{c^2(\xi)-U^2(\xi)}
 \left(\din_{x_i}^{\xi}\!\dfrac{r_0(\eta)}{U(\eta)}\,
 \dfrac{[R_0+\Pi(\eta)]^2}{c^2(\eta)-U^2(\eta)}\,\dd\eta\right)\,\dd\xi \\ \ \\
{} = I_{2i}- \dfrac{R_0+c^2_2}{2\mu_2}\left[I_{1i}\ln{|x-x_2|} -
 \dfrac{R^2_0-c^4_2}{4\mu_2c^2_2}\,\ln^2{|x-x_2|}\right] + O(|x-x_2|),
 \ea
 \]
 where $I_{2i} = $\;const. Therefore, when $x\to x_2$, the solution is:
 \be
 \ba{l}
 r(x) = \dfrac{R_0-c^2_2}{R_0+c^2_2}\left[1 + \dfrac{\ri\,\om}{\mu_2}\,\ln{|x-x_2|} - \dfrac{\om^2}{2\mu^2_2}\,\ln^2{|x-x_2|} + \dots\right]
 \\ \ \\ \phantom{wwa}
 {} - \dfrac{2\ri\,\om\, c^2_2I_{1i}}{(R_0+c^2_2)^2} \left[1 + \dfrac{\ri\,\om}{\mu_2}\,
 \ln{|x-x_2|} + \dots\right] - \dfrac{4\om^2c^2_2I_{2i}}{(R_0+c^2_2)^3}+O(|x-x_2|).
 \ea
 \label{r(x)}
 \ee

On the other hand, setting in Eq. (\ref{rx0}) $x_0 = x_2$ and $\mu = \mu_2$, we obtain for $|x-x_2|\ll 1$ and $\om|\ln|x-x_2||\ll 1$:
 \[
 \ba{l}
 r(x)=r_{sl}(x) + B^{(\pm)}_0|x-x_2|^{\ri\,\om/\mu_2}\Bl[1+O(|x-x_2|)\Br]
  \\ \ \\ \phantom{wwl}
 {} = r_{sl}(x)+B^{(\pm)}_0\left[1+\dfrac{\ri\,\om}{\mu_2}\,\ln{|x-x_2|} - \dfrac{\om^2}{2\mu^2_2}\,\ln^2{|x-x_2|} + \dots\right]\Bl[1 + O(|x-x_2|)\Br].
 \ea
 \]
Matching of this solution with solution (\ref{r(x)}) shows that when $x\to x_2\pm 0$ the solution should be as this:
 \be
  \label{rx2}
 r(x) = r_{sl}(x)+\left[\dfrac{R^{(\pm)}_0-c^2_2}{R^{(\pm)}_0+c^2_2} - \dfrac{2\,\ri\,\om\, c^2_2I^{(\pm)}_{1i}}{(R_0+c^2_2)^2} + O(\om^2)\right]|x-x_2|^{\ri\,\om/\mu_2}\Bl[1+O(|x-x_2|)\Br],
 \ee
where $r_{sl}(x)$ is determined by the series (\ref{Slow}) with $x_0 = x_2$. It was taken into account that $R_0$ and $I_{1i}$ can be different on the left and right of point $x_2$, because according to the boundary conditions (\ref{BC}), they are calculated through the integration of Eq. (\ref{r}) in the different flow regions. Therefore, if $r_0(x_2\pm 0)\ne 0$ (i.e. $R^{(\pm)}_0\ne c^2_2$), then the reflection coefficient and transmission ratios are:
$$
 |r(x)| \to \left|\dfrac{R^{(\pm)}_0-c^2_2}
 {R^{(\pm)}_0+c^2_2}\right|^{1/2} = O(1),
$$
 \be
 \label{T2}
 K_2\approx\left[1-\left|\dfrac{R^{(-)}_0-c^2_2}{R^{(-)}_0+c^2_2} \right|^2 \right]^{-1/2}, \quad 
 K_3\approx\left[1-\left|\dfrac{R^{(+)}_0-c^2_2}{R^{(+)}_0+c^2_2} \right|^2 \right]^{1/2}.
 \ee
Otherwise, $|r(x)|$ is of the order of $O(\om^2)$.

It should be emphasized that for $x \to x_2 \pm 0$, the quantity 
\[
r_0(x_2\pm 0)=\dfrac{R^{(\pm)}_0-c^2_2}{R^{(\pm)}_0+c^2_2}
\]
determined by Eq. (\ref{r0}) does not describe the limiting value of $r(x)$ (or its main part), but, as seen from Eq. (\ref{rx2}), {\it the amplitude of the rapidly oscillating part of the solution}. Note that oscillations are concentrate in the exponentially narrow neighborhood of the point $x_2$,
\be
 |x-x_2|=O(\re^{-1/\om}).
 \label{dx2}
 \ee
This allows us, firstly, to consider $r_0(x_2 \pm 0)$ as the main part of the intermediate asymptotic of the solution, and, secondly, greatly complicates detection of oscillations in the numerical solution even in the case of not very small $\om$ (see Section \ref{sec:5} and Fig.~\ref{f02}).

Based on these results, we will consider wave propagation within each section of the flow. Let's start from the left interval, $-\infty < x < x_1$, where $U(x) < c(x)$. Given that $a(x) \ne 0$ only in the interval $\sigma = (x_-, \, x_1)$, we consider two options. The first option is that $x_1 - x_- = O(1)$ and $a(x) = O(1)$ in the entire interval $\sigma$. Then, taking into account the boundary condition (\ref{BC}), we obtain (see equations (\ref{r01}), (\ref{mr}), and (\ref{K123})):
 \be
 r_- \equiv r(x_-)=\dfrac{1-\Pi_-}{1+\Pi_-}+O(\om) \quad \mbox{\rm and}
 \quad  K_1 = \sqrt{1-\left(\dfrac{1-\Pi_-}{1+\Pi_-}\right)^2} + O(\om),\ \
 \Pi_-=\Pi(x_-).
 \label{r-}
 \ee

The subcritical flow simulates the ``ordinary'' space-time, with the quite natural constancy of the wave velocity, $c \equiv 1$. Then $\Pi_- = U(-\infty) < 1$ and there is necessarily a reflected wave; the transmission rate is $K_1 < 1$. The reflectionless propagation with $\Pi_- = 1$ is possible only in a more complex model, where $c(x) > 1$ for $x \to -\infty$ and decreases to 1 as we approach the critical point $x_1$ (BH horizon) -- see, for example, Fig. \ref{f01}(c).

 In the second option, we compose the interval $\sigma$ from two intervals of the length $O(1)$ each, $\sigma_1 = (x_-, x_a)$ and $\sigma_2 = (x_b, x_1)$. In these intervals function $a(x) = O(1)$; the intervals are separated by the insert $\sigma_0 = (x_a, x_b)$ where $a(x) \equiv 0$. Then, Eq. (\ref{r01}) gives $r_0(x_b) = (1 -\Pi_0)/(1 + \Pi_0)$, where $\Pi_0 = \Pi(x_a) = \Pi(x_b)$. As $a \equiv 0$ within the insert, therefore, $|r(x)| = $ const, and only the phase difference of waves $\theta(x)$ varies with $x$; however, this variation affects the transmission rate. Indeed, in the case of the `phase inverting insert', when
\be
 \theta(x_a)-\theta(x_b)=\om\!\din_{x_a}^{x_b}\!\al(x)\,\dd x=(2n+1)\pi,
 \label{L}
 \ee
 where $n$ is natureal, $r_0(x_a) = -r_0(x_b)$, so that in accordance with Eq. (\ref{r0}), $R_0 = \Pi^2_0$ and
 \be
 r_0(x_-)=\dfrac{\Pi^2_0-\Pi_-}{\Pi^2_0+\Pi_-}, \qquad
 K_1 = \sqrt{1-\left(\dfrac{\Pi^2_0-\Pi_-}{\Pi^2_0+\Pi_-}\right)^2}.
 \label{r-2}
 \ee
As a result, the waves reflected in the intervals $\sigma_1 $ and $\sigma_2$ cancel out each other upon the condition $\Pi^2_0 = \Pi_-$, which can be easily implemented even in traditional flow models with $c \equiv 1$ (see Fig. \ref{f01}(a)), if in the insert section $U(x) = U_a = U_b = \sqrt{U(-\infty)}$ (in this case, $U(-\infty) < U_a < 1$). This equality sets the position of the anti-reflective insert; its length is determined by Eq. (\ref{L}) and is very long, since it is proportional to $\om^{-1}$.

In conclusion, we note that wave propagation in the interval $x_2 < x < \infty$ does not differ qualitatively from that just described. Indeed, since $a(x) = 0$ for $x > x_+$, then $\Pi_+ \equiv \Pi(x_+) = \Pi(+\infty)$ and integration of Eq. (\ref{r}) starts at $x = x_+$ with $r(x_+) = r(+\infty) = 0$. If $x_+ - x_2 = O(1)$ and $a(x) = O(1)$, then
\be
 r_0(x_2+0)=\dfrac{\Pi_+ -\Pi_2}{\Pi_+ +\Pi_2}, \qquad
 K_3 = \sqrt{1-\left(\dfrac{\Pi_+ -\Pi_2}{\Pi_+ +\Pi_2}\right)^2} + O(\om),
 \label{r-3}
 \ee
 and the condition of suppression of the reflected wave, $\Pi_+ = \Pi_2$, is satisfied if $c(+\infty) > c_2$, i.e. requires an increase in the wave speed downstream from the critical point $x_2$ (see Fig. \ref{f01}(c)). But it is possible to suppress the reflected wave using the above-described `anti-reflective optics effect', i.e. with the help of inverting insert with $a(x) = 0$ in such point $x = x_a$ where $\Pi^2(x_a) = \Pi_2\Pi_+$.
 
 Let us now turn to the region $x_1 < x < x_2$ where the flow is supercritical. If $x_2 - x_1 = O(1)$ and $a(x) = O(1)$, then Eq. (\ref{r01}) gives:
 \be
 r_0(x_2-0)=\dfrac{1-c^2_2}{1+c^2_2}, \qquad
 K_2=\left[1-\left(\dfrac{1-c^2_2}{1+c^2_2}\right)^2\right]^{-1/2} + O(\om) = \dfrac{1+c^2_2}{2c_2} + O(\om).
 \label{x2}
 \ee
 In the case when the wave speeds at the ends of the region are the same ($c_2 = c_1 = 1$), $r_0(x_2-0) = 0$, and $K_2 = 1 + O(\om^2)$ is almost not
 differs from one. Again, one can significantly increase the transmission ratio using an inverting insert. Indeed, since $\Pi_1 = \Pi_2 = 1$, there is a point $x = x_m$ at which $\Pi(x)$ reaches extreme (maximum or minimum) value of $\Pi_m$. Here $a(x)$ changes sign, and the growth of $|r(x)|^2$ is replaced by a decrease, which leads to a decrease in the amplitudes of waves of both positive and negative energy. Changing at this point (more precisely, on the inverting insert with $a = 0$) the sign of $\cos{\theta}$ (and with it the $r$ sign), we will continue to grow $|r(x)|^2$,
 \be
 r_0(x_m-0)=\dfrac{1-\Pi_m}{1+\Pi_m}\ \longrightarrow\ r_0(x_m+0)=-r_0(x_m-0)\
 \longrightarrow\ r_0(x_2-0)=\dfrac{\Pi^2_m-1}{\Pi^2_m+1}\,,
 \label{x2-2}
 \ee
 and amplitudes of both waves. As the result, $K_2 \approx (\Pi^2_m + 1)/(2 \Pi_m) > 1$, and to the greater extent, the more strongly $\Pi_m$ differs from 1 (to either way, since $K_2 (\Pi_m) \approx K_2(\Pi^{-1}_m)$).

  If $c_2 \ne 1$, then $K_2 > 1$ and without an inverting insert, and, in addition, $K_2(c_2) = K_2(c_2^{- 1})$ up to $O(\om)$. In other words, amplification of a wave of positive energy due to its transformation into a wave of negative energy is promoted by both a decrease and an increase in
  speed of waves in the supercritical section of the flow.
\subsection{The high-frequency limit, $\om\gg 1$}
\label{ssec:4.3}
Let us turn now to high frequency waves, $\om \gg 1$. This limiting case can be still consistent with the shallow-water approximation which requires in the dimensional variables that $\omega \gg ck$, whereas $kh \ll 1$ and $\sigma \ll \rho g h^2/3$, where $\sigma$ is the surface tension. As follows from the analysis carried out in Section \ref{sec:3}, in the vicinity of the critical point $x_1$, function $r(x)$ is defined by the series (\ref{Slow}) the coefficients of which have the order of $O(\om^{-1})$. Therefore, it is natural to assume that solution to Eq. (\ref{r}) has the same order. Setting $r(x) = \ri P(x)/\om$ where $|P| = O(1)$, but $|\dd P/\dd x| = O(\om)$, we get the equation:
\[
 \dfrac{1}{\om}\,\dfrac{\dd P}{\dd x} = -\ri\,\al(x)P + \ri\, a(x)\left(1 + \dfrac{P^2}{\om^2}\right).
 \]
 Neglecting the term $O(\om^{-2})$ in this equation, we arrive at the linear non-homogeneous equation whose solution subject to the boundary conditions (\ref{BC}) is:
 \be
 P(x) = \ri\,\om\, e^{-\ri\,\om\Psi(x)}\din_{x_0}^{x} a(\xi)\, e^{\ri\,\om\Psi(\xi)}\,\dd\xi\,, \quad \mbox{where} \quad
 \Psi(x) = \int\!\al(x)\,\dd x,
 \label{P}
 \ee
 and $x_0 = x_1$ when $x<x_2$ and $x_0 = x_+$ when $x > x_2$.
 
 In Eq. (\ref{P}), when calculating the integral of a rapidly oscillating
 function, we take into account that $\dd\Psi/\dd  x= \al(x)$ has no zeros on the real axis and therefore, there are no stationary phase points, but $\al(x)$ has
 poles at $x=x_{1,2}$. Hence, the main contribution to the integral comes from
 a neighborhood of the upper limit of integration (see, for example, \cite{Olver}), then we have:
 \be
 \din_{x_0}^{x} a(\xi)\,e^{\ri\,\om\Psi(\xi)}\,\dd\xi\, \sim
 \dfrac{-\ri\,a(x)}{\om\,\al(x)}\,e^{\ri\,\om\Psi(x)} = O(\om^{-1}).
 \label{IP}
 \ee
 We see that indeed  $P(x)=O(1)$ in general. However, 
 $P(x) = o(1)$ when $x\to x_2\pm 0$ because of presence of $\al(x)$ in the denominator. Thus, in the high-frequency limit $K_1$ differs from unity by $O(\om^{-2})$ whereas $K_2$ and $K_3$ differ from unity even less, only by $o(\om^{- 2})$. Therefore, both the wave reflections in the subcritical regions of the flow and wave amplification in the supercritical region diminish.
 \section{Wave propagation in the WH--BH duct model}
 \label{sec:5}
 Let us consider now a duct with a supercritical flow ($U(x) > c(x)$)
 in the outer domains (left: $-\infty < x < x_1$), and right: ($x_2 < x < +\infty$), and subcritical in the inner domain ($x_1<x<x_2$). here we will use dimensionless variables such that (cf. Eq.~(\ref{scale})):
 \be
 U_2 = c_2 = 1, \qquad \mu_2 = U'_2 - c'_2 = 1.
 \label{scale2}
 \ee 
 
 In this case, the physical statement of the problem is not so evident as for the BH--WH duct considered in Section~\ref{sec:4}; we will present the detail discussion of this delicate issue in Section~\ref{sec:7}. Here we study a restricted problem of wave propagation only in the rightmost domain where $x>x_1$. Let us start again with the BH horizon (at the point $x=x_2$ rather than at $x=x_1$!) because at this point both the reflected wave and NEW vanish. Assume that the co-current propagating wave arriving from the domain where $x < x_1$ has the amplitude $D(x_2) = 1$. Then, in the domain $x_1 < x < \infty$ (cf. Eqs~(\ref{D1}) and (\ref{K})) we have:
 \bea
  \label{D2}
 D(x) & = & \exp\left[-\!\din_{x_2}^{x}a(x')\,r(x')\,\dd x'\right], \\
  \label{K2}
 K(x) & = & \!\exp\left[\!-\!\din_{x_2}^{x} a(x')\RE\Bl(r(x')\Br)\,\dd x'\right]\!\equiv\!\exp\left[-\!\din_{x_2}^{x} a(x')|r(x')|\cos\theta(x')\,\dd x'\right],
 \eea
 and the integral transmission ratios of sub- and supercritical domains are (cf. Eq.~(\ref{K123})):
 \be
 K_2=\sqrt{1-|r(x_1)|^2}\le 1, \quad
 K_3=\Bl[1-|r(+\infty)|^2\Br]^{-1/2}\ge 1, \quad
 K(+\infty)=K_2 \cdot K_3.
 \label{K23}
 \ee
 It should be borne in mind that the oscillating component of $r(x)$
 develops with the distance from $x_2$ rather than from $x_1$.
 
 Scaling as per Eq.~(\ref{scale2}) yields $\Pi_2 \equiv c_2U_2 = 1$, therefore, the scattering of low frequency waves in the subcritical domain is approximately described by Eq.~(\ref{r01}). Then, the transmission ratio in this domain is:
 \be
 K_2=\sqrt{1-\left(\dfrac{1-c^2_1}{1+c^2_1}\right)^2} + O(\om) =
 \dfrac{2c_1}{1+c^2_1} + O(\om).
 \label{K2-2}
 \ee
 From this formula, it follows (bearing in mind a small contribution of the last term $O(\omega)$) that $K_2$ is only slightly less than unity if the wave velocities in the ending points of the domain are the same, $c_1 = c_2 = 1$, but $K_2$ decreases when $c_1$
 deviates from $c_2$ in the either side. It should be also noted that $K_2(c_1) \approx K_2(1/c_1)$.
 
 In the high-frequency limit, one can demonstrated by analogy with Section~\ref{ssec:4.2} that $K_2 \to 1-0$ for any $c_1$. As a consequence of this, when $c_1 = 1$, $K_2(\om)$ goes to unity in the both limiting cases, when $\omega \to 0$ or $\omega \to \infty$, and inside this frequency range, it has at least one minimum. If $c_1 \ne 1$, then $K_2(\om)$ grows from
 $K_2(0) < 1$ up to 1, but not necessarily monotonically (see right panel in Fig.~\ref{f10}). 
 
 In the right outer domain $x > x_2$, the transformation of a positive-energy low-frequency wave into a negative energy wave is described by the same equation (\ref{r01}), and the transmission ratio is:
 \be
 K_3=\left[1-\left(\dfrac{\Pi(+\infty)-1}{\Pi(+\infty)+1}\right)^2
 \right]^{-1/2} + O(\om) =
 \dfrac{\Pi(+\infty)+1}{2\sqrt{\Pi(+\infty)}} + O(\om).
 \label{K3-2}
 \ee
 Asymptotically $K_3 \sim \sqrt{\Pi(+\infty)}/2$ increases with $\Pi(+\infty) \equiv c(+\infty)U(+\infty)$. However, when the frequency increases $K_3$ approaches unity from the top.
 
 In the conclusion of this section we note that in Ref. \cite{Euve17} it was demonstrated experimentally that analogue traversable and bidirectional wormholes can exist in dispersive hydrodynamics when the surface tension effect is taken into consideration. Then the capillary wavelength plays the role of a Planckian scale below which long gravity waves are transformed
into short capillary waves that are able to move with the ``superluminal'' speeds. Whereas the results obtained are not applicable to putative astrophysical wormholes {\it per se}, they are of interest from the hydrodynamic point of view.

\section{Results of numerical calculations}
\label{sec:6}

\subsection{BH--WH duct}
 \label{ssec:6.1}
The analysis presented above shows that the most interesting scenarios of wave propagation occur in the middle, supercritical, flow region where a positive energy wave is amplified due to the coupling with the negative energy wave; then the transmission coefficient can be noticeably greater than one. Numerical calculations were performed mainly for this region. 
Recall that at the far end of the region where $x_2 - x \sim \exp(-\om^{-1})$, the function $r(x)$ rapidly oscillates; therefore, the spatial resolution of the oscillating solution in the numerical calculations can be performed without extra complications only for relatively high $\om$.
Below we consider separately two particular cases when the wave speed is (i) the same at the ending points of the supercritical interval and (ii) when it is different.
\subsubsection{Currents with the equal velocities at the ending points}
\label{ssec:6.2}
Bearing in mind the conditions (\ref{scale}), we take for calculations a simple flow model with $c_1=c_2=1$ (see Fig. \ref{f01}(a)):
\be
c(x)\equiv 1, \qquad U(x)=1+x(2d-x)/(2d), \quad x_1=0,\ \ x_2=2d.
 \label{M1}
 \ee

\begin{figure}[h!]
\centerline{\includegraphics[width=15cm]{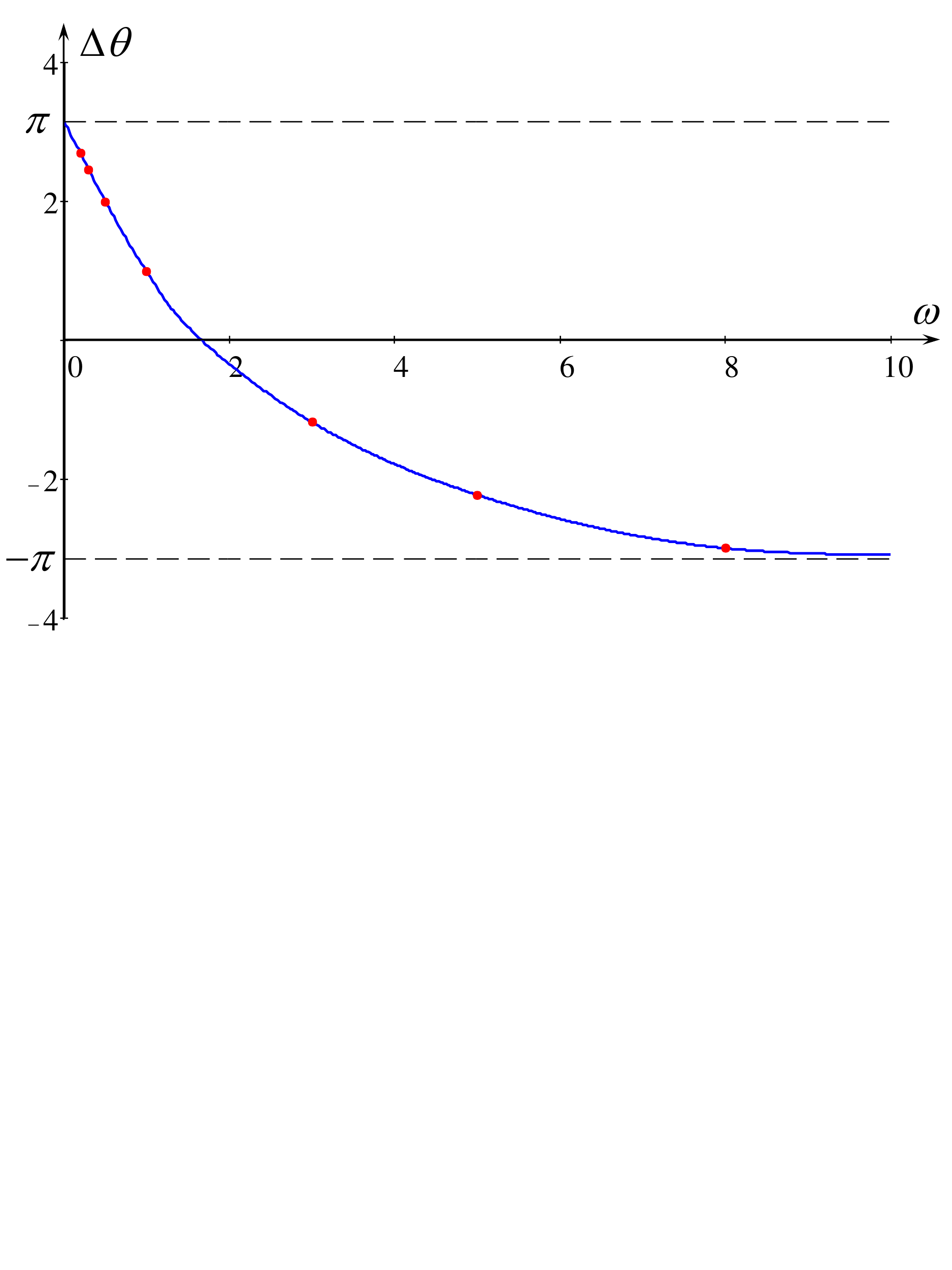}} %
\vspace*{-10 cm}%
\caption{(color online) \protect\footnotesize
Optimal phase shift versus frequency. Dots show the values used in the numerical calculations.}
\label{f15}
\end{figure}

 For such a flow, the Mach number attains its maximum,
 ${\cal M} = 1 + d/2$, at the midpoint of the supercritical interval, $x_m = d$, and function $a(x)$ changes its sign. The phase-shifting insert at this point is simulated by changing of the phase $\theta = \arg(r)$ by the required value $\Delta\theta$. Note that in the limit $\om \to 0$, the most effective insert is inverting one, with $\Delta \theta = \pm \pi$, we have considered it above.
 However, when the frequency increases, the phase $\theta$ changes more and more significantly in that region where $a(x) \ne 0$, therefore the optimal phase shift that maximizes the transmission ratio $K_2$, depends on the frequency. Specific values of $\Delta \theta(\om)$ used in our calculations for the flow with ${\cal M} = 4$ ($d = 6$) are given in Table \ref{tab1} (see also Fig. \ref{f15}). 
 \begin{table}[h]
 \caption{Data for the optimal phase shift versus frequency used in the numerical calculations.}
 \begin{center}
  \begin{tabular}{|c||c|c|c|c|c|c|c|}        \hline
 $\om$          & 0.2   & 0.3   & 0.5   & 1     & 3      & 5      & 8
  \\ \hline
 \phantom{.}$\Delta\theta$ \phantom{.}&\phantom{.} 2.688 \phantom{.}&\phantom{.} 2.443 \phantom{.}&\phantom{.} 1.990 \phantom{.}&\phantom{.} 0.977 \phantom{.}&\phantom{.} $-1.187$ \phantom{.}&\phantom{.} $-2.234$ \phantom{.}&\phantom{.} $-3.002$ \phantom{.}  \\ \hline
  \end{tabular}
 \end{center}
 \label{tab1}
 \end{table}

\begin{figure}[t!]
\vspace*{-2cm}%
\centerline{\includegraphics[width=17cm]{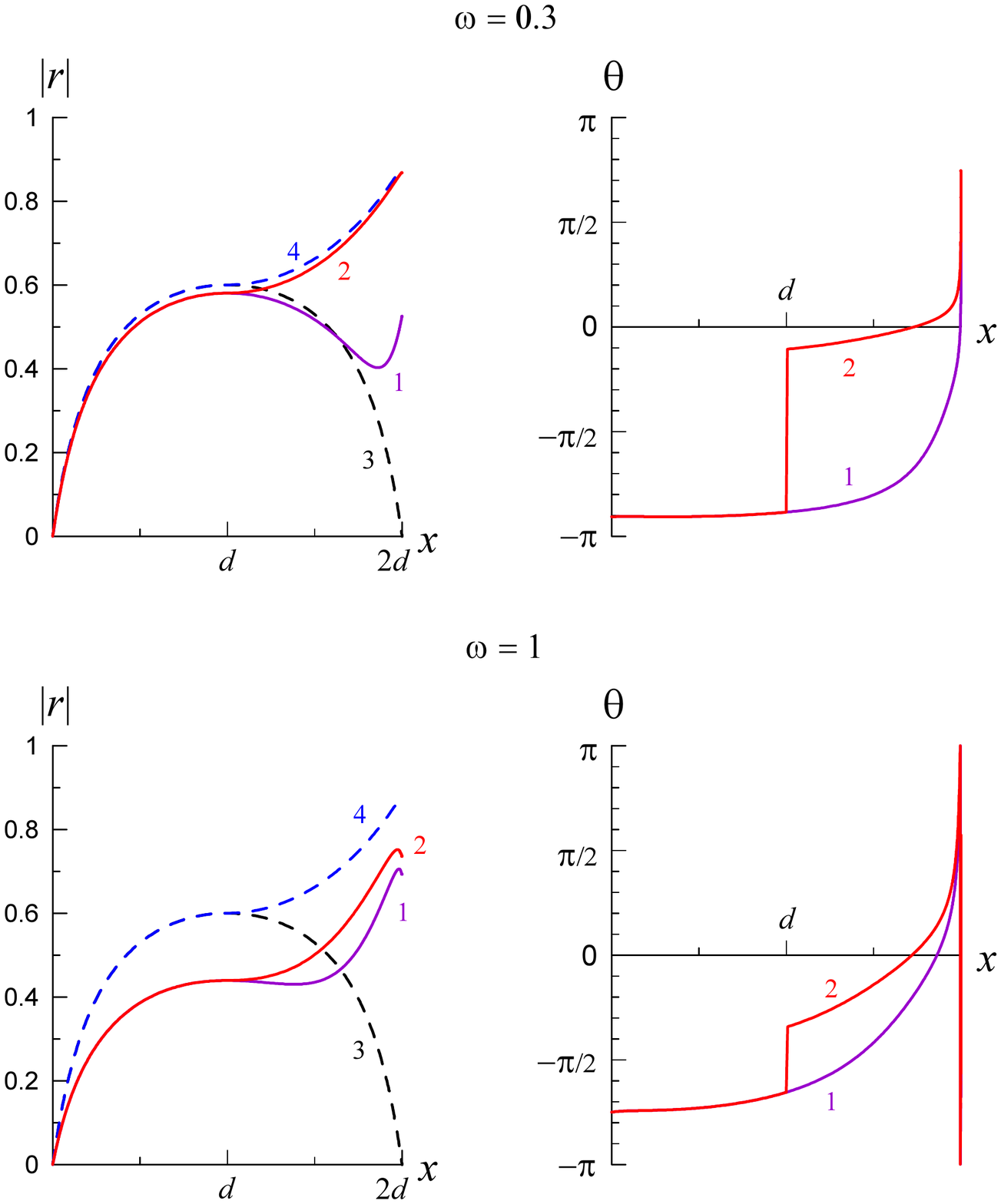}} %
\vspace*{-4.5cm}%
\caption{(Color online) \protect\footnotesize
The reflection coefficient (left panels) and function $\theta(x)$ (right panel) as functions of $x$ for three values of frequency $\om\le 1$ in the flow model (\ref{M1}) with ${\cal M} = 4$:\ \ line 1 -- NI;\ \ line 2 -- OI. Dashed lines correspond to the reference case with $\om = 0$ and $\Delta\theta = \pi$: \ \ line 3 -- NI; \ \ line 4 -- OI.}
\label{f02}
\end{figure}
Figure \ref{f02} shows the reflection coefficient $|r(x)|$ that describes the transformation of the co-current propagating wave into the negative energy wave, as well as $\theta(x) \equiv \arg(r)$ for the frequencies $\omega \le 1$, both with the optimal phase-shifting insert (OI), and without it (NI).
The upper part of Fig. \ref{f02} shows that even when the frequency is not very small, the ``zero'' approximation $|r_0(x|$ as defined by the Eq. (\ref{r0}) (and with phase inversion, if any) describes pretty well the behavior of function $|r(x)|$, except for the neighborhood of the point $x = x_2 = 2d$ in the NI version when there is no insert. The difference between the curves is caused by missing in $r_0(x)$, but accumulated in $r(x)$ (due to $\om \ne 0$) a rapidly oscillating component, although its amplitude is significantly less than in the OI version.
\begin{figure}[b!]
\vspace*{-2.cm}%
\centerline{\includegraphics[width=17cm]{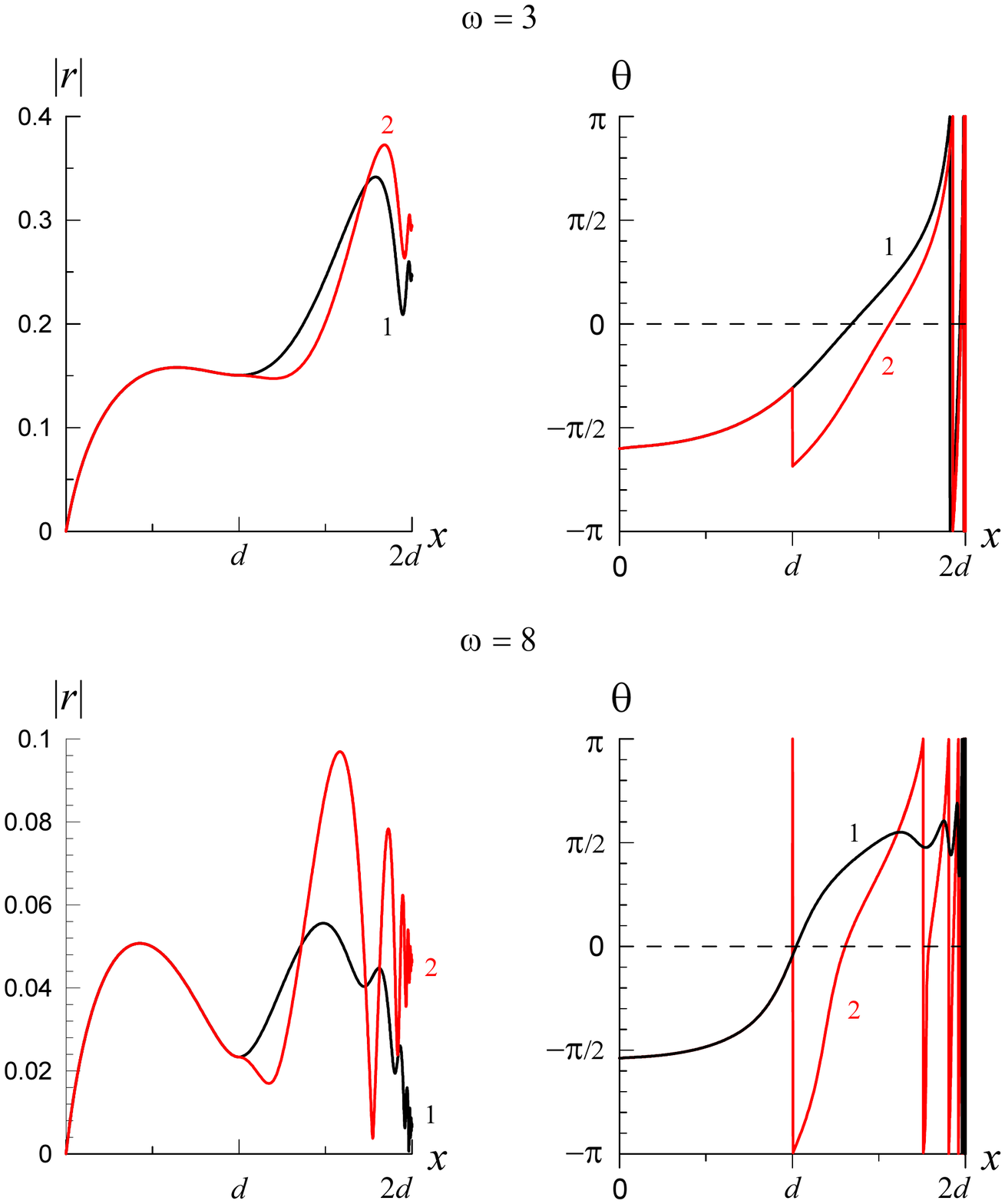}} %
\vspace*{-3.5cm}%
\caption{(Color online) \protect\footnotesize
The reflection coefficient (left panels) and function $\theta(x)$ (right panel) as functions of $x$ for two values of the frequency $\om > 1$ in the flow model (\ref{M1}) with ${\cal M} = 4$: \ \ line 1 -- NI; \ \ line 2 -- OI.}
\label{f03}
\end{figure}
 For $\om = 1$, the downstream change in $\theta$ manifests faster, therefore $|r(x)|$ and $|r_0(x)|$ differ notably throughout the entire supercritical region, in both versions of the flow, with the insert (OI-version) and without the insert (NI-version). Note that when $x\to x_2 = 2d$, the reflection coefficients (which are equal here to the amplitudes of fast oscillations of the functions $r(x)$) corresponding to the different flow versions, approach each other (cf. lines 1 and 2 in Fig. \ref{f02}) and slightly decrease compared with what was for $\om = 0.3$ in the OI-version. Moreover, we draw attention to the fluctuation of $|r(x)|$ near the point $x_2$, which became noticeable in the result of expansion of the region of oscillations.

 Figure \ref{f03} shows the reflection coefficient $|r(x)|$ and function $\theta(x)$ for $\om > 1$ for the both flow models, OI (with insert) and NI (no insert). Here the region of oscillations is even wider, and oscillations of $|r(x)|$ are seen in a notably greater range of $x$. When $x \to x_2$, these oscillations, as expected, decay, and $|r(x)|$ approaches a finite limit which is equal to the amplitude of oscillations of the function $r(x)$. Comparison of graphs for $\om = 1, \ 3, \ 8$ confirms the conclusion made at the end of Section \ref{sec:4} that $|r(x)|$ decreases as $\om^{-1}$, and even faster in the vicinity of $x_2$.

\begin{figure}[b!]
\vspace*{-2.5cm}%
\centerline{\includegraphics[width=18cm]{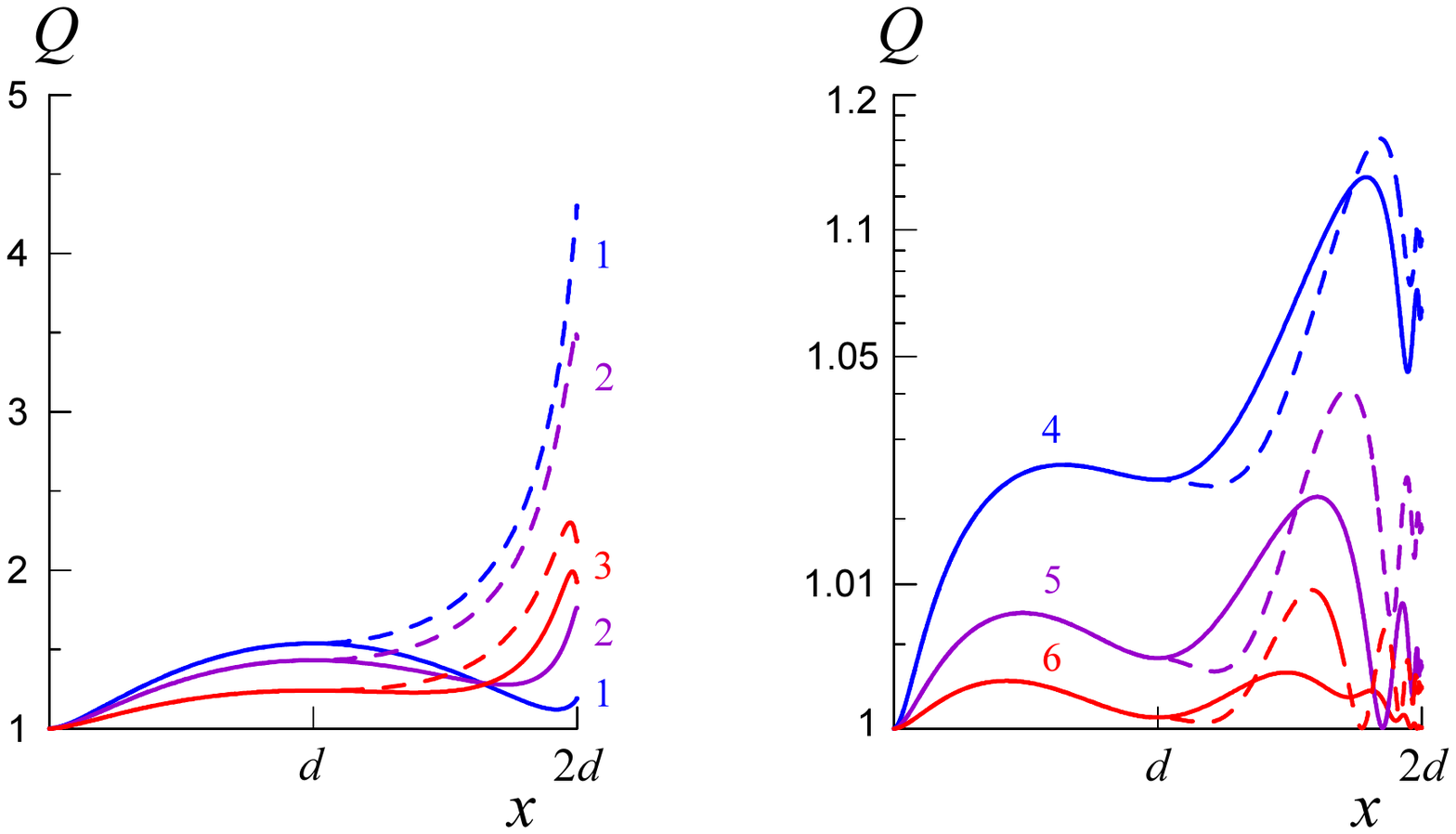}} %
\vspace*{-14.0cm}%
\caption{(Color online) \protect\footnotesize
The gain coefficient as the function of $x$ for the several particular frequencies: \ \ 1 -- $\om=0.2$, \ \ 2 -- $\om = 0.5$, \ \ 3 -- $\om = 1$, \ \ 4 -- $\om = 3$, \ \ 5 -- $\om = 5$, \ \ 6 -- $\om = 8$. Solid lines pertain to the NI model, dashed lines -- to the OI model.}
\label{f04}
\end{figure}

 In the course of propagation from $x_1$ to $x_2$, waves of positive and negative energies interact such that the total energy flux (\ref{En2}) conserves. As the result, their amplitudes synchronously increase or decrease in according with the change in their phase difference. Quantitative measure of wave interaction is the gain of the positive energy wave:
 \be
 Q(x)=\dfrac{1}{1-|r(x)|^2};
 \label{Q}
 \ee
 this quantity is shown in Fig. \ref{f04}. According to Eq. (\ref{mr}), the growth is replaced by the decrease when the sign of either $a(x)$ or $\cos{\theta}$ changes.

At low frequencies ($\om < 1$) in the OI model of the flow, the change of the function $a(x)$ sign is compensated to a large extent by the phase jump, therefore $Q(x)$ grows monotonically and begins slightly oscillate only near the point $x_2$ (see curves 3).
  As the frequency increases, $\theta(x)$ varies more and more rapidly, and $Q(x)$ acquires more and more distinct oscillatory character whereas its value becomes closer and closer to unity.

\begin{figure}[h!]
\vspace*{-4.cm}%
\centerline{\hspace{5cm}\includegraphics[width=17cm]{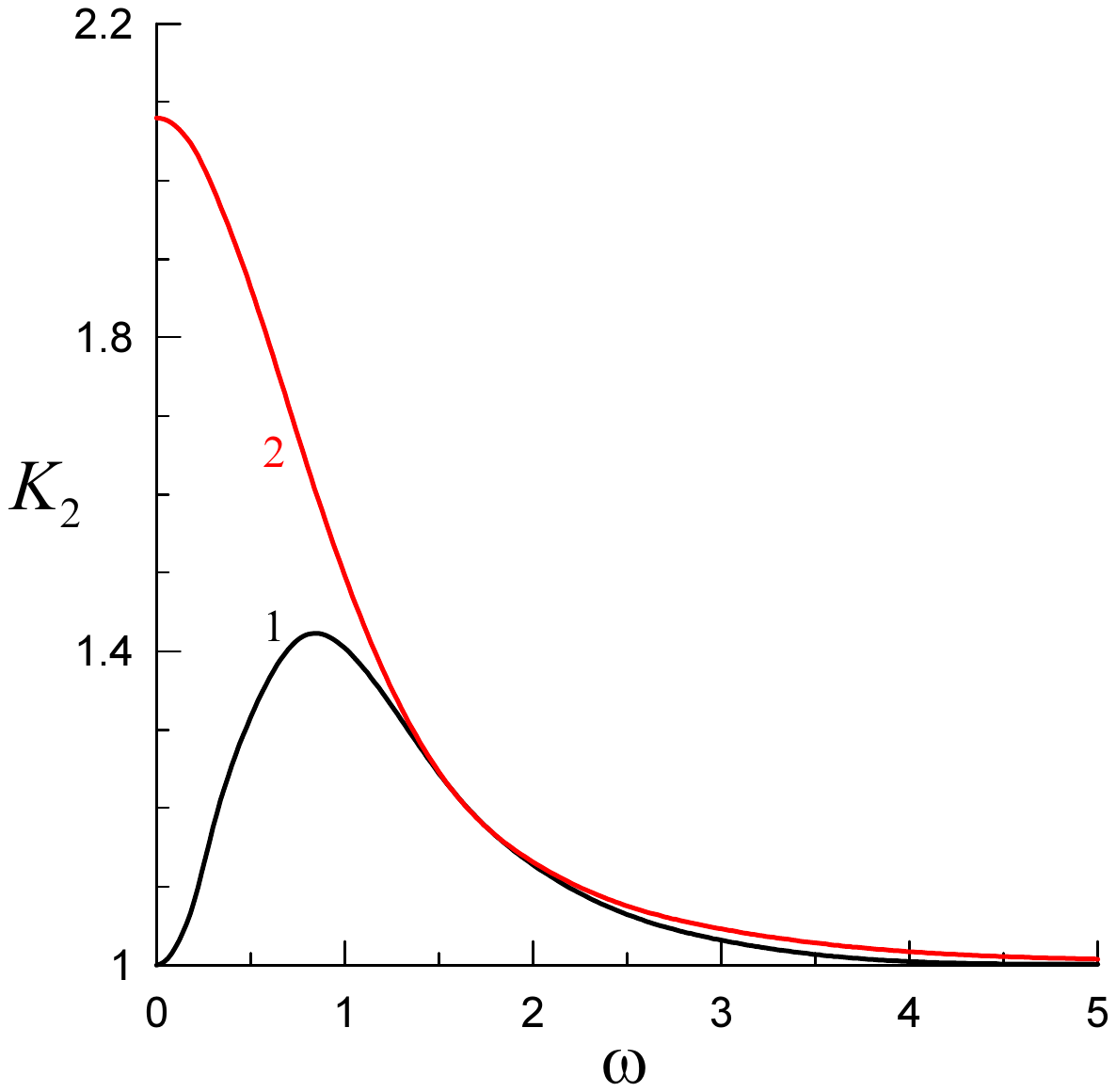}} %
\vspace*{-11.5cm}%
\caption{(Color online) \protect\footnotesize
The transmission ratio $K_2$ as functions of frequency $\omega$ for the NI (line 1) and OI (line 2) models.}
\label{f04K}
\end{figure}

  Figure \ref{f04K} shows the dependences of the transmission ratio $K_2$ of the supercritical domain as the functions of frequency $\omega$ for the NI (line 1) and OI (line 2) models. As one can see from the comparison of lines 1 and 2, both models, with the optimal phase-shifting insert (OI) and without it (NI), provide approximately the same transmission ratio for $\omega > 1$, whereas they differ in the low-frequency domain. Whereas the OI model provides monotonic increase of $K_2$ when $\omega \to 0$, in the NI model, $K_2$ has a maximum at $\omega = 1$, and then goes to zero when $\omega \to 0$. Even in the NI model without any insert, the transmission ratio is notably greater than one, what can provide the laser effect of wave amplification in the active zone. 

\begin{figure}[b!]
\vspace*{-3.5cm}%
\centerline{\includegraphics[width=17.5cm]{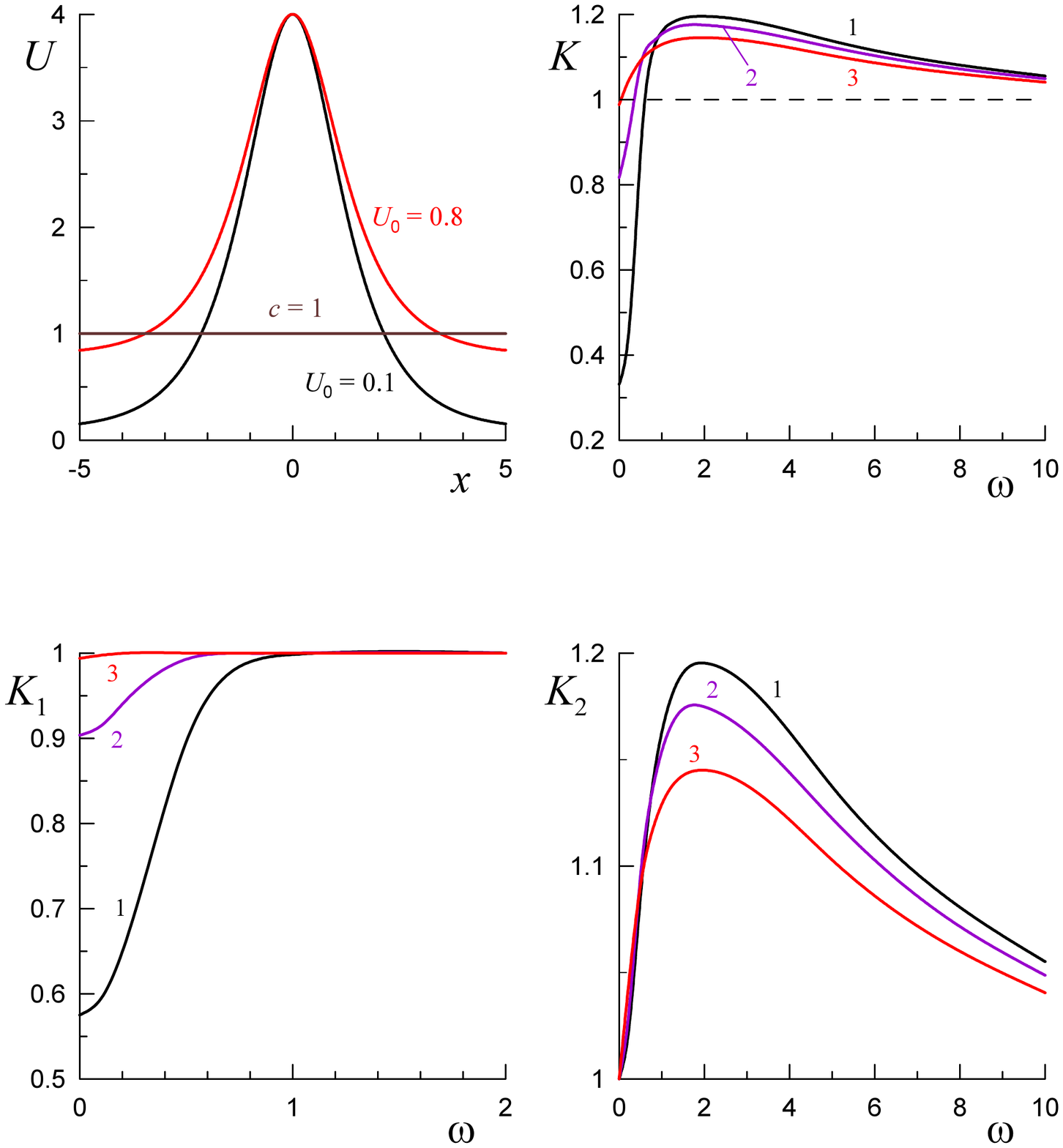}} %
\vspace*{-4.5cm}%
\caption{(Color online) \protect\footnotesize
Top panels: left -- velocity profiles (\ref{MU}) for $U_0 = 0.1$ and $U_0 = 0.8$, and right -- the total transmission ratio $K = K_1 K_2 K_3$ as function of frequency $\omega$. Bottom panels: left -- transmission ratios of the left ($K_1$) domain and right -- for the  middle ($K_2$) domain as functions of frequency $\omega$;\ \ 1 -- $U_0 = 0.1$,\ \ 2 -- $U_0 = 0.4$,\ \ 3 -- $U_0 = 0.8$.}
\label{f14}
\end{figure}

The amplitude gain in the supercritical domain may be drastically reduced by wave reflection in the left and right subcritical domains. To gain a better insight into the effect of domain competition, we have calculated the wave transmission through all three domains in the bell-shaped velocity profile (see Fig.~\ref{f01}a) with following dependences:
  \be
  c(x) \equiv 1, \quad {\rm and} \quad
  U(x) = U_0 + (M - U_0)\sech{x}.
  \label{MU}
  \ee
Calculations were performed for the fixed Mach number ${\cal M} = 4$ and with various velocities at the infinity $U_0 \in (0,\,1)$. The results obtained are presented in Fig. \ref{f14}.
As one can see, the supercritical flow with any $U_0$ acts as a broadband amplifier (see the top right panel in Fig.~\ref{f14}). This can be explained as follows: the wave reflection plays a noticeable role only at low frequencies, whereas the transmission ratio $K_2$ responsible for the wave amplification decreases with $\om$ rather slowly. It should be noted in passing that due to the flow symmetry, the transmission ratios in the subcritical domains of the flow are equal, $K_1 = K_3$, so that the total transmission ratio is $K = K_1^2 K_2$. And finally, when comparing $K_2(\om)$ shown in Fig. \ref{f14} with that presented in Fig. \ref{f04K} by curve 1, it should be born in mind that in calculations presented in Fig. \ref{f14} the scaling (\ref{scale}) was not fulfilled. Due to this, the dimensionless frequencies of the maximal amplification are different.
\subsubsection{Currents with the different velocities at the ending points}
\label{ssec:6.3}
In this section, we consider currents in which wave velocity $c(x)$ monotonically increases or decreases downstream from $c_1 = 1$ to $c_2$. For calculations, we take a modification of the flow model (\ref{M1}) without a phase-shifting insert (see Fig. \ref{f01}(b)),
\be
 c(x)=1+\dfrac{c_2-1}{2d}\,x, \qquad U(x)=c(x)+\dfrac{x(2d-x)}{2d},
 \quad x_1=0,\ \ x_2=2d.
 \label{M2}
 \ee
In such a flow, the maximum Mach number is attained at $x = 2d/(1 + \sqrt{c_2})$:
\be
 {\cal M}=1+\dfrac{2d}{(1+\sqrt{c_2})^2}\,,  \label{Md1}
 \ee
so that for the given Mach number we have:
\be
 d =  \dfrac{{\cal M}-1}{2}\Bl(1+\sqrt{c_2}\Br)^2.
 \label{Md2}
 \ee
\begin{figure}[b!]
\vspace*{-2.5cm}%
\centerline{\includegraphics[width=17.5cm]{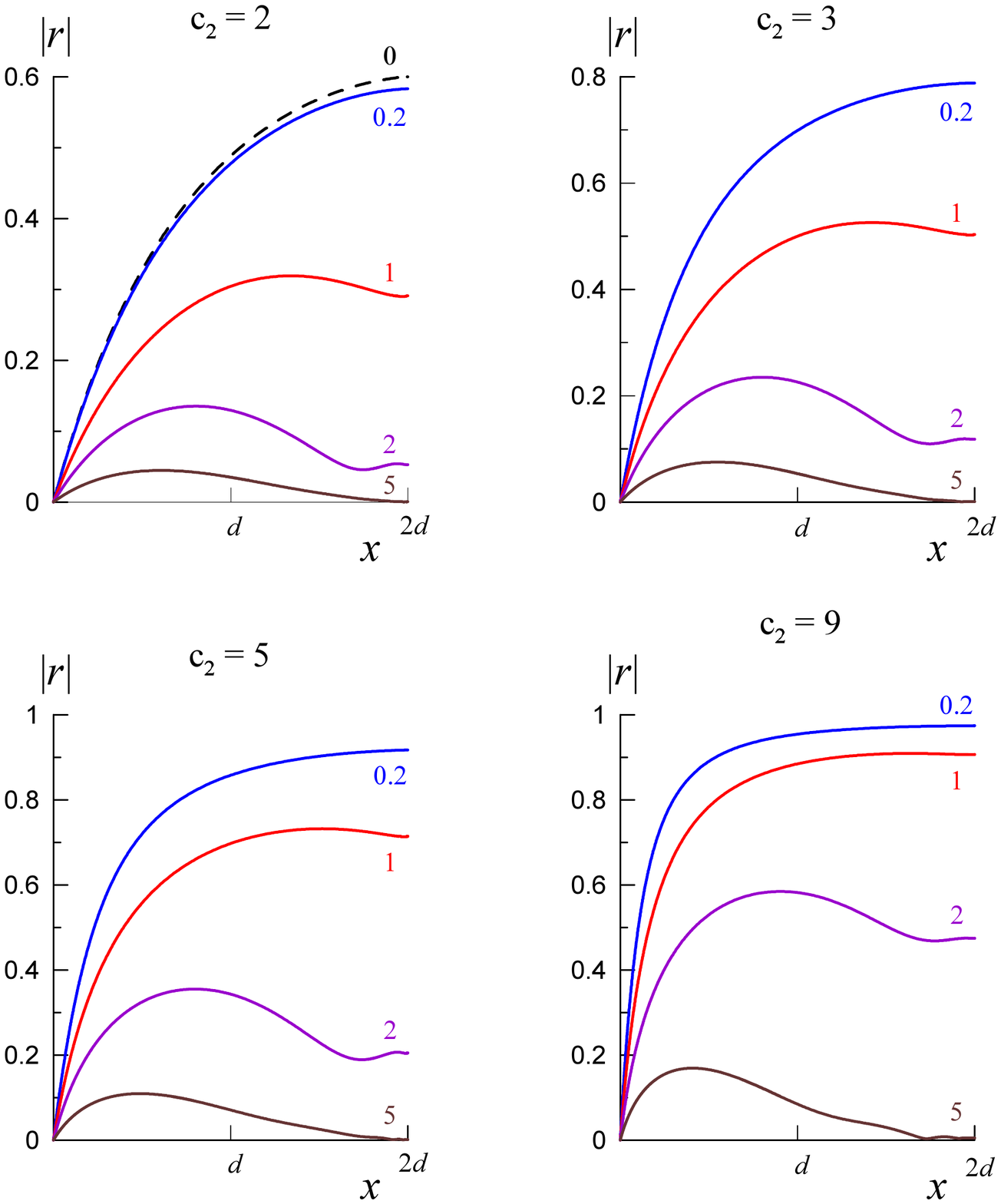}} %
\vspace*{-4.5cm}%
\caption{(Color online) \protect\footnotesize
Dependence of $|r(x)|$ in the flows with increasing wave velocity $c(x)$ for different values of $c_2$ and frequencies shown by the numbers next to the curves. Dotted line pertains to $\om = 0$.}
\label{f05}
\end{figure}

Let us assume that in the entire region function $\Pi(x) = c(x)U(x)$
varies monotonically, and $a(x)$ does not change its sign. This limits the range of the flow parameters, so that
 \be
 d\le |c_2-1|, \qquad 1<{\cal M}\le\left\{
 \ba{ll}
 3-\dfrac{4}{1+\sqrt{c_2}}\,, \quad \mbox{if} \quad c_2>1,\\
 \dfrac{4}{1+\sqrt{c_2}}-1\,,  \quad \mbox{if} \quad  c_2<1.
 \ea
 \right.
 \label{dM}
 \ee
\begin{figure}[t!]
\vspace*{-2.5cm}%
\centerline{\includegraphics[width=18cm]{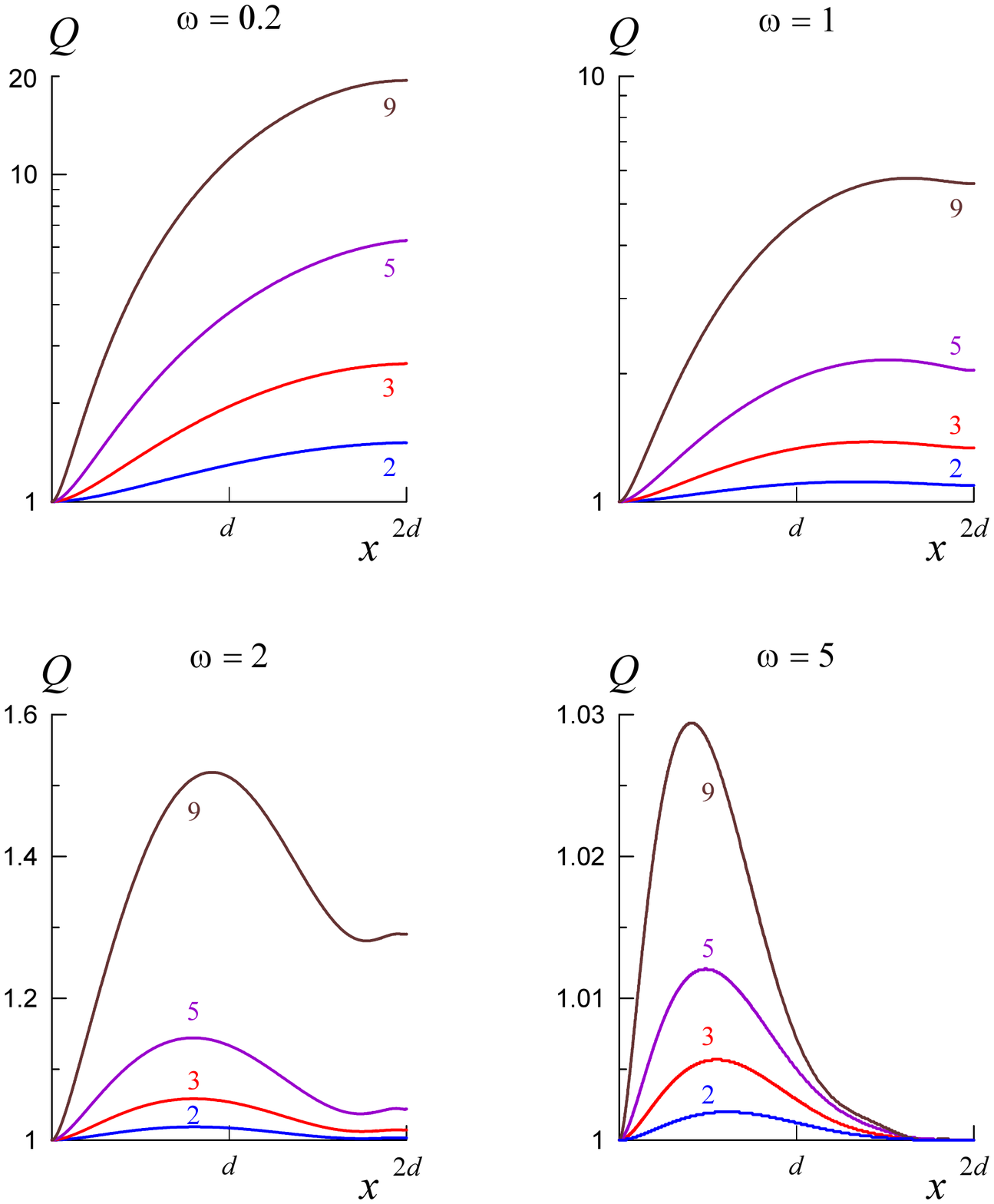}} %
\vspace*{-4.5cm}%
\caption{(Color online) \protect\footnotesize 
The gain coefficient in the flows with increasing wave velocity $c(x)$ for the different frequencies and ending values of $c_2$ shown by numbers next to the curves).}
\label{f06}
\end{figure}
The parameters $c_2$ and ${\cal M}$ used for calculations are shown in Table \ref {tab2}. For a given $c_2$, the Mach number ${\cal M}$ was chosen close to the maximum value, and then the parameter $d$ was calculated using Eq. (\ref{Md2}).
 \begin{table}[!h]
 \caption{The Mach numbers used in the numerical calculations with the different values of $c_2$.}
 \begin{center}
  \begin{tabular}{|l||c|c|c|c||c|c|c|c|c|}        \hline
 $\phantom{.}c_2\phantom{.}$ & 2 & 3 & 5 & 9 & $2/3$ & 0.5 & $1/3$ &
 $1/6$ & 0.1  \\ \hline
 $\phantom{.}{\cal M}\phantom{.}$  &\phantom{.} 1.3 \phantom{.}&\phantom{.} 1.5 \phantom{.}&\phantom{.} 1.6 \phantom{.}&\phantom{.} 2 \phantom{.}&\phantom{.} 1.2 \phantom{.}&\phantom{.} 1.3 \phantom{.}&\phantom{.} 1.5 \phantom{.}&\phantom{.} 1.8 \phantom{.}&\phantom{.} 2 \phantom{.} \\ \hline
  \end{tabular}
 \end{center}
 \label{tab2}
 \end{table}
\begin{figure}[h!]
\vspace*{-2.5cm}%
\centerline{\includegraphics[width=16.5cm]{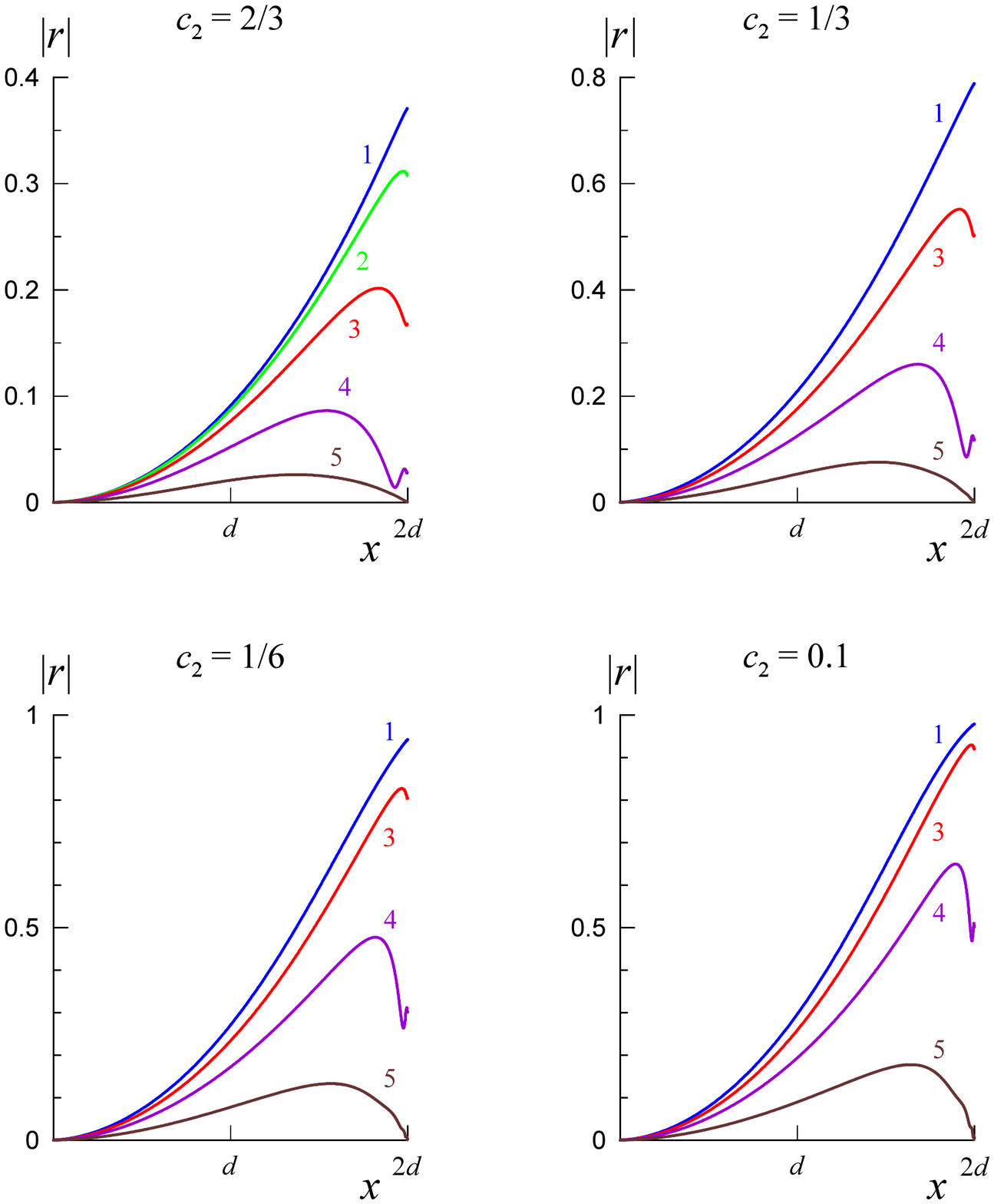}} %
\vspace*{-4.0cm}%
\caption{\protect\footnotesize (Color online)
 Dependence of $|r(x)|$ in the flows with the decreasing wave velocity $c(x)$ for the different values of $c_2$ and frequencies: \ \ 1 -- $\om = 0.2$, \ \ 2 -- $\om = 0.5$, \ \ 3 -- $\om = 1$, \ \ 4 -- $\om = 2$, \ \ 5 -- $ \om = 5$.
}
\label{f07}
\end{figure}
For the flows with the increasing velocity $c(x)$, Fig. \ref{f05} shows the dependence $|r(x)|$ for different $c_2$ and $\om$. It is clearly seen that the transformation into the negative energy wave increases with increasing of $c_2$, but falls abruptly when the frequency growths. This is also exhibits by the graphs of the gain coefficient $Q$ for the same set of parameters (see Fig. \ref{f06}).

\begin{figure}[h!]
\vspace*{-2.0cm}%
\centerline{\includegraphics[width=17.5cm]{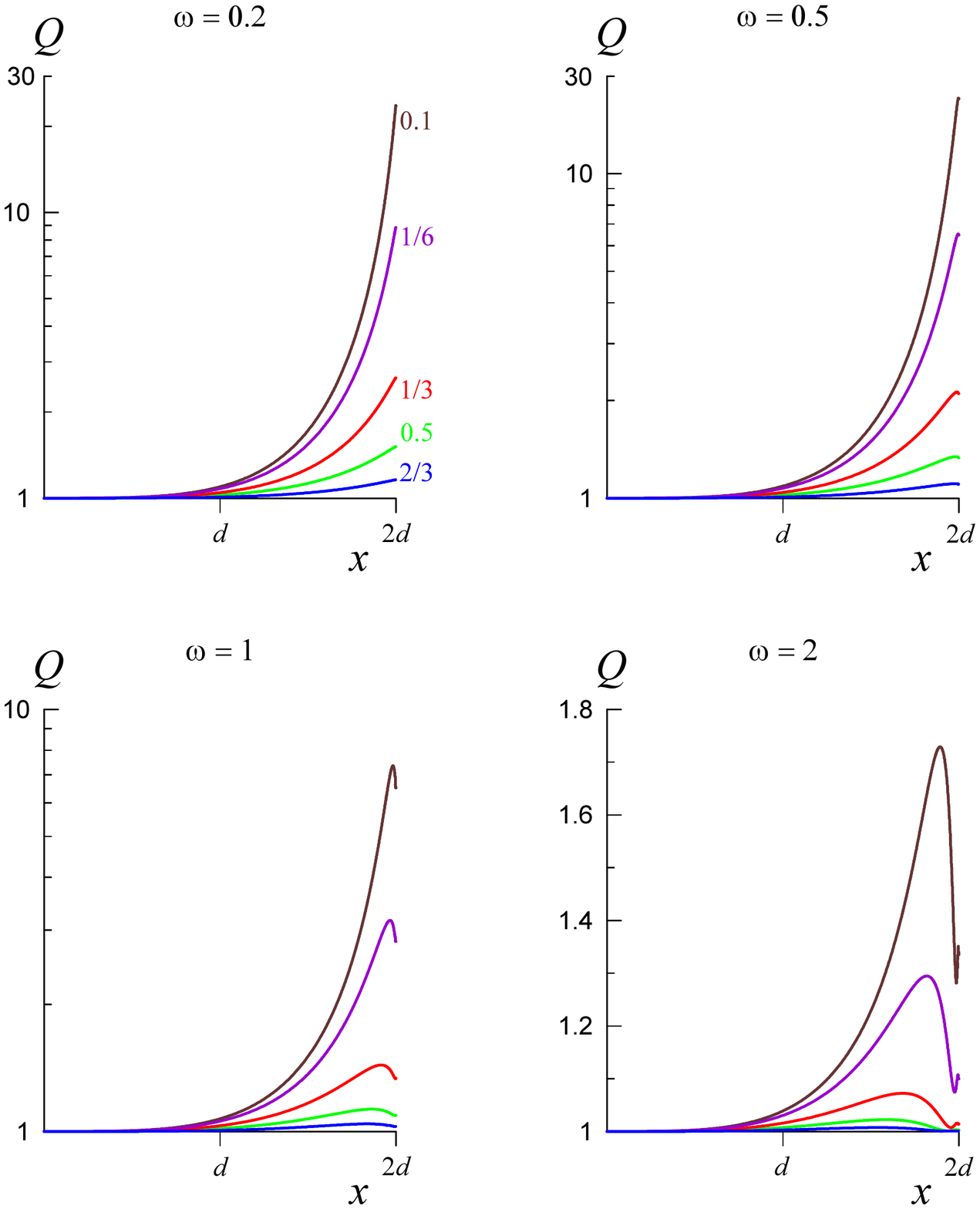}} %
\vspace*{-4.0cm}%
\caption{\protect\footnotesize (Color online) The gain coefficient in the flows with decreasing wave velocity $c(x)$ for the different frequencies and ending values of $c_2$ shown by numbers next to the curves in the upper left panel.}
\label{f08}
\end{figure}

Similar dependences for the flows with the decreasing function $c(x)$ are shown in Figs. \ref{f07} and \ref{f08}. They demonstrate the strengthening of the wave transformation with the decrease of the ending value $c_2$ and its sharp weakening with the increase of $\om$. The oscillatory nature of the transformation near $x_2$ is seen even more clearly than for $c_2 > 1$.

\begin{figure}[h!]
\vspace*{-5.5cm}%
\centerline{\includegraphics[width=18cm]{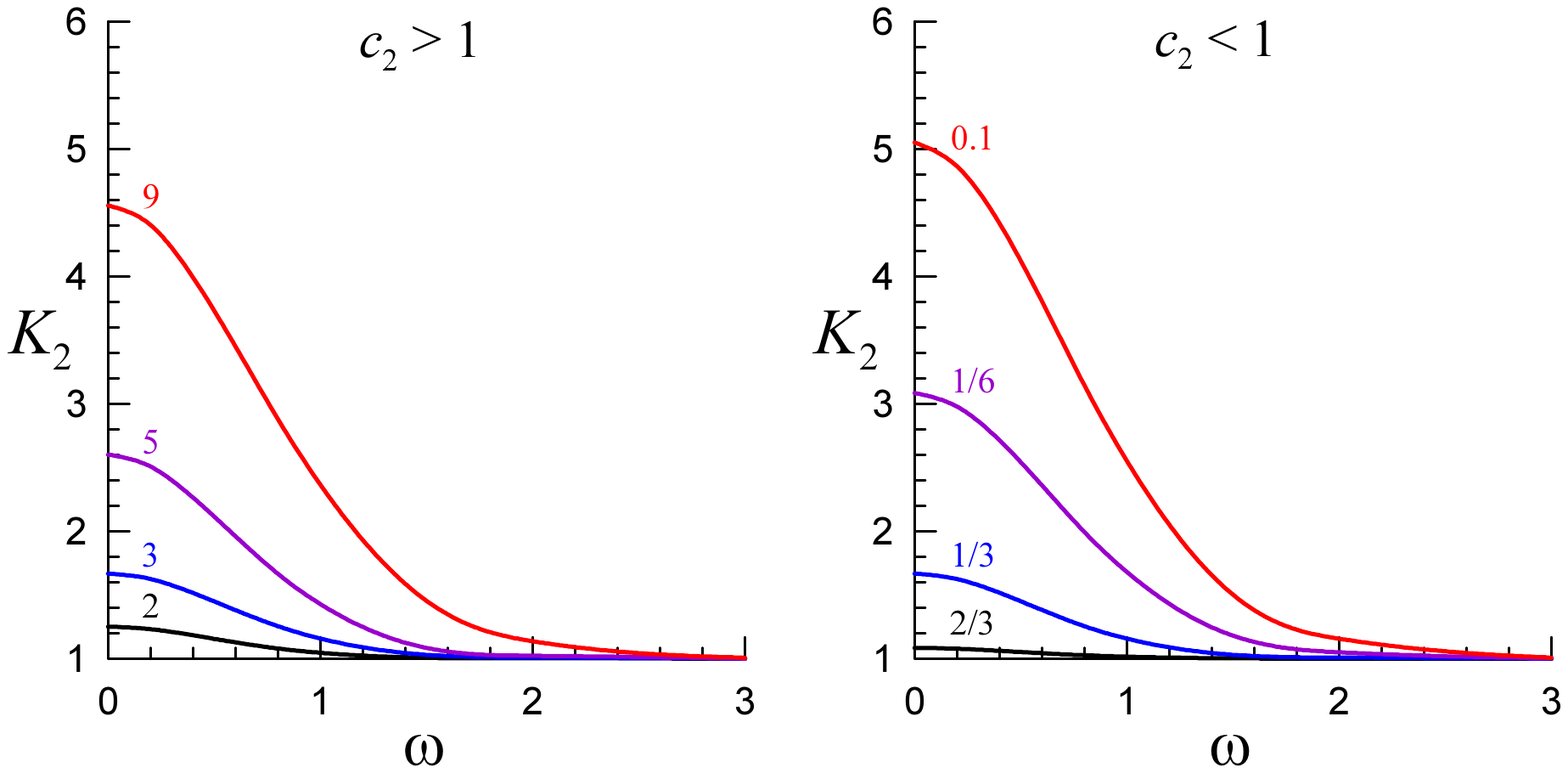}} %
\vspace*{-12.0cm}%
\caption{(Color online) \protect\footnotesize
The dependences of the transmission ratio $K_2$ of the inner domain as functions of frequency $\omega$ for the different values of $c_2 > 1$ (left panel) and $c_2 < 1$ (right panel). Numbers next to curves show the values of $c_2$.}
\label{f06-8}
\end{figure}

Figure \ref{f06-8} shows the transmission ratio $K_2$ of the inner domain as function of frequency $\omega$ for the different values of $c_2 > 1$ (left panel) and $c_2 < 1$ (right panel). As one can see from the figure, the transmission ratio in the both panel monotonically decreases from some maximal value $K_2(\omega = 0)$, which depends on $c_2$, to zero.
 \subsection{WH--BH duct}
 \label{ssec:6.4}
 For this type of ducts, the inner domain, $x_1 < x <x_2$, is subcritical; the corresponding transmission ratio is less than one, $K_2(\omega) \le 1$. The frequency dependence of $K_2(\omega)$ is determined by the particular profiles $c(x)$ and $U(x)$ (see Section \ref{ssec:4.3}). For the numerical calculations, we chose a family of flows with the linear $c(x)$ and quadratic $U(x)$ profiles
 (cf. Eqs. (\ref{M1}) and (\ref{M2})),
 \be
 c(x) = 1+\dfrac{(c_1-1)(2d-x)}{2d}, \qquad U(x) = c(x)-\dfrac{x(2d-x)}{2d}.
 \label{M3}
 \ee
 Here $c_1=c(0)$, and $d$ is chosen such that $U(x)>0$ in the entire interval $x_1=0\le x\le x_2=2d$.
 
\begin{figure}[h!]
\vspace*{-4.0cm}%
\centerline{\includegraphics[width=17cm]{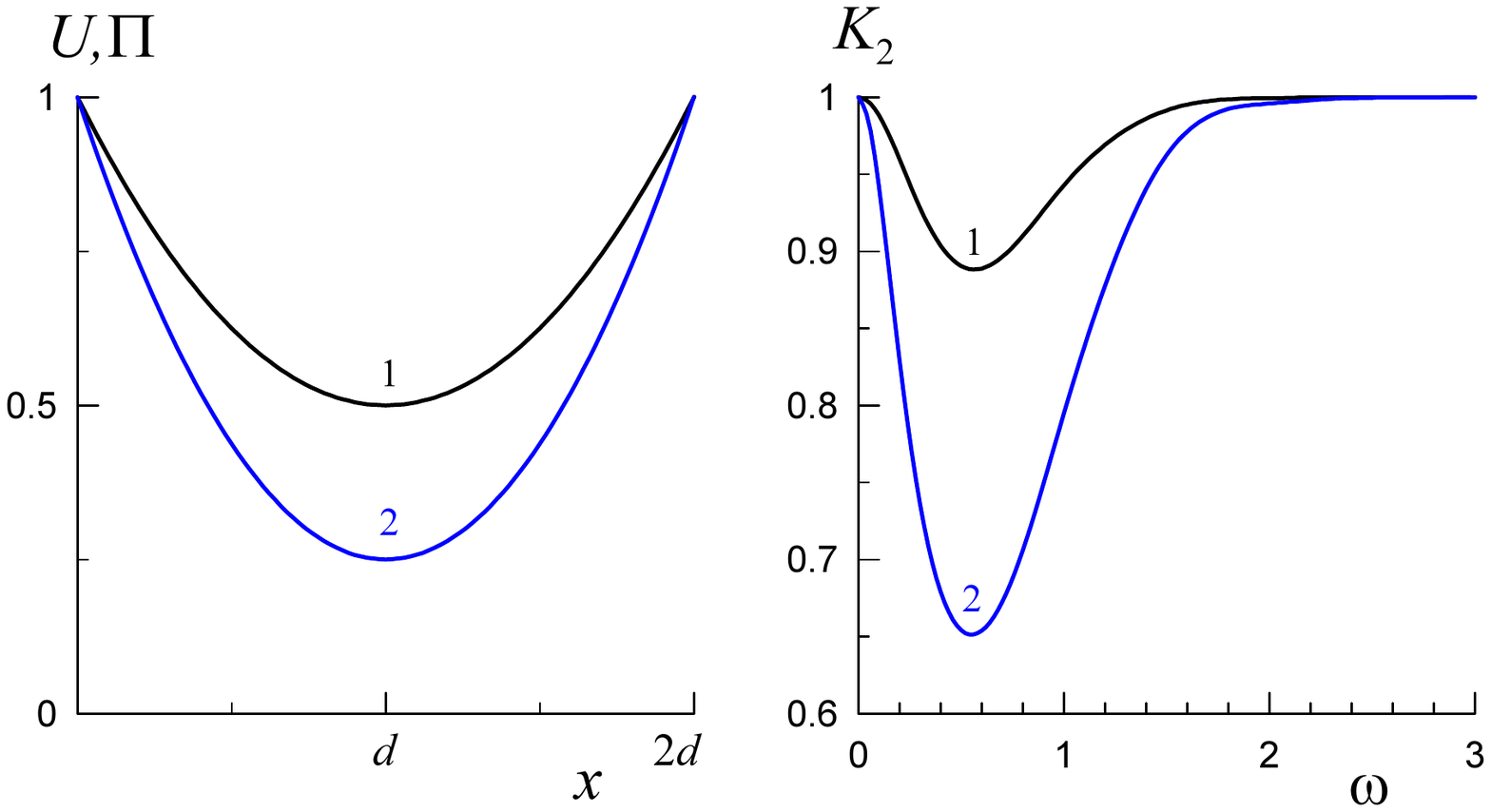}} %
\vspace*{-11.5cm}%
\caption{\protect\footnotesize (Color online) Graphics of the flow as per Eq. (\ref{M3}) with $c_1 \equiv 1$.
Left panel: the flow velocity $U(x) \equiv \Pi(x)$ as the function of $x$. Right panel: the frequency dependence of the transmission ratio $K_2(\omega)$.  Lines 1 are plotted for $d = 1$ with $U_{min} = 0.5$; lines 2 -- for $d = 1.5$ with $U_{min} = 0.25$.}
\label{f09}
\end{figure}

\begin{figure}[h!]
\vspace*{-4.5cm}%
\centerline{\includegraphics[width=16.5cm]{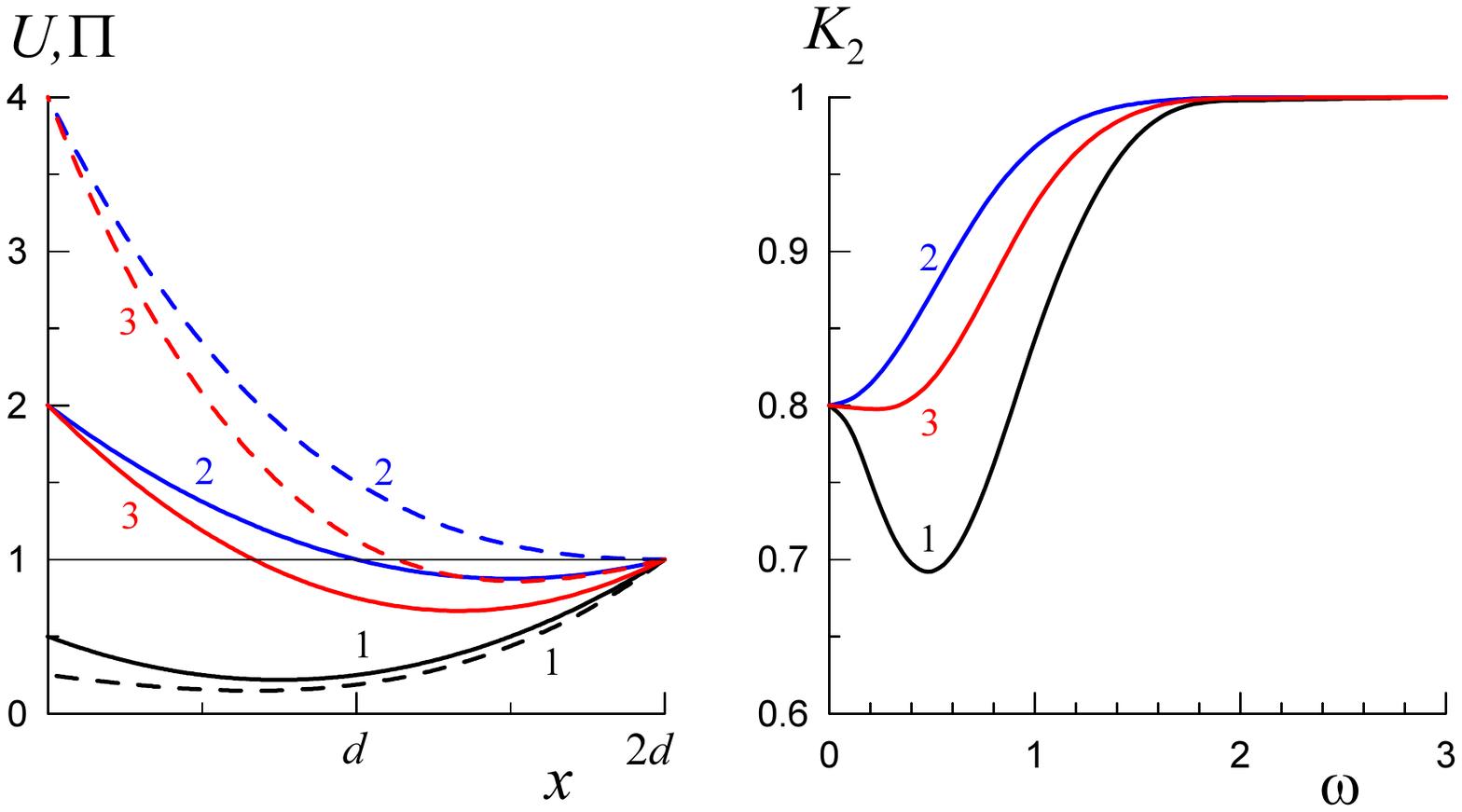}} %
\vspace*{-11cm}%
\caption{\protect\footnotesize (Color online) Graphics of the flow as per Eq. (\ref{M3}) with $c_1 \ne 1$. Left panel: the flow velocity $U(x)$ (solid lines) and $\Pi(x)$ (dashed lines) as functions of $x$. Right panel: the frequency dependence of the transmission ratio $K_2(\omega)$. Lines 1 are plotted for $c_1 = 0.5$, $d = 1$; lines 2 -- for $c_1 = 2$, $d = 1$; lines 3 -- for $c_1 = 2$, $d = 1.5$.}
\label{f10}
\end{figure}

  If $c_1=c_2=1$, then $c(x)\equiv 1$, and $\Pi(x)=c(x)U(x)\equiv U(x)$. In this case, the transmission ratio $K_2(\om)\to 1$ when the frequency goes either to zero or to infinity, and its minimum decreases with the decreasing of $U_{min}=(1-d/2)$ (see Fig.~\ref{f09}). If, however, $c_1 \ne 1$, then $K_2(0) < 1$ (see Eq.~(\ref{K2-2})) and
 $K_2(\om)$ increases to 1 (not necessarily monotonically) when frequency grows (see right panel in Fig.~\ref{f10}).

\begin{figure}[h!]
\vspace*{-4.5cm}%
\centerline{\includegraphics[width=16.5cm]{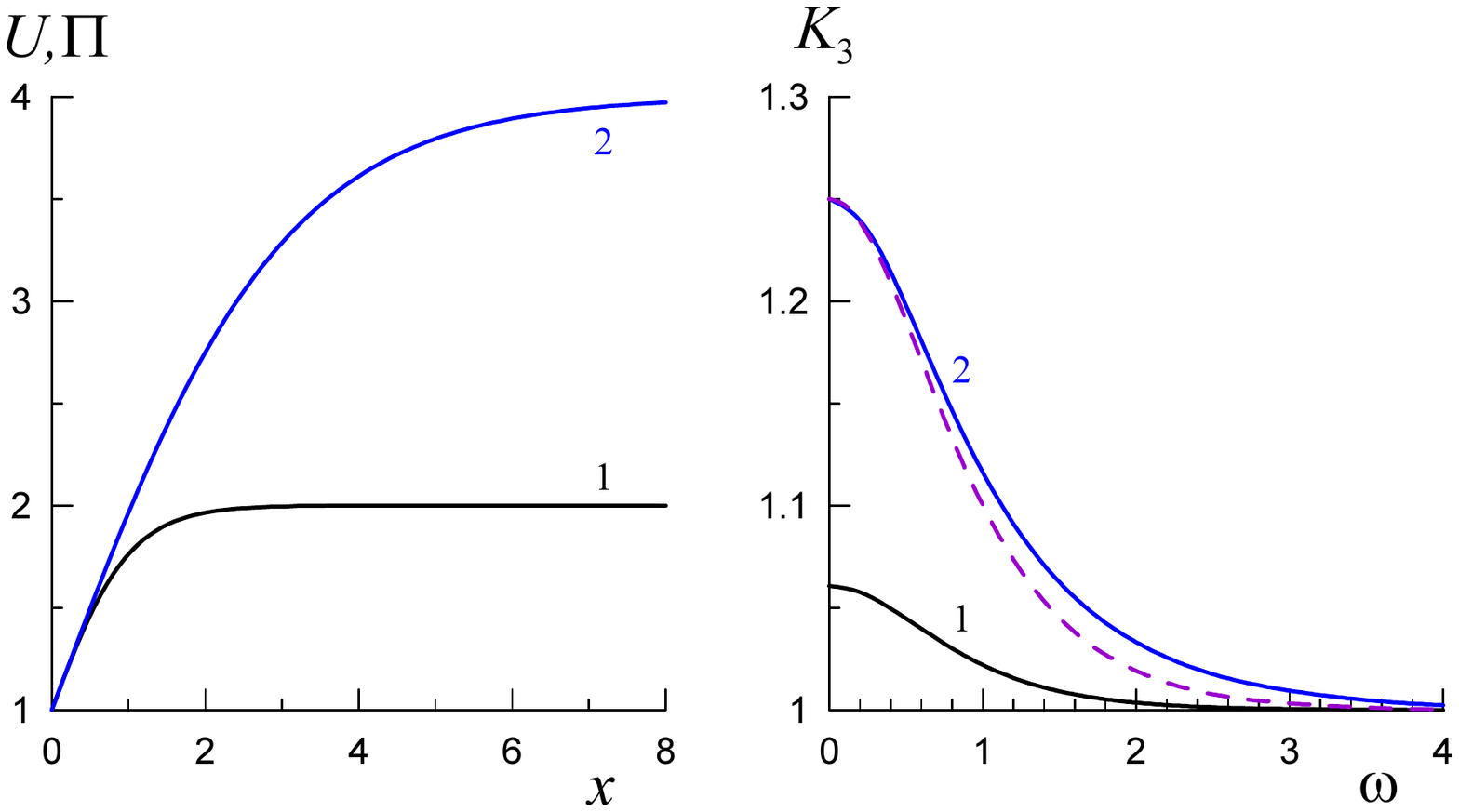}} %
\vspace*{-11cm}%
\caption{\protect\footnotesize (Color online) The flow (\ref{M4}), $c(x) \equiv
1$. Left panel: flow velocity $U \equiv \Pi$ vs $x$. Right panel: frequency
dependence of the transmission ratio $K_3$. \ \ 1 -- $D = 1$ ($U_{max} = 2$), 
\ \ 2 -- $D = 3$ ($U_{max} = 4$). For comparison, the dashed line presents the
curve 2 plotted in the right panel of Fig.~\ref{f12}.}

\label{f11}
\end{figure}

The positive energy wave is amplified after passing into the supercritical domain $x > x_2$ due to the interaction with the NEW \cite{FabrStep}, therefore, the transmission ratio $K_3 \ge 1$. For the calculations, we choose $c(x)$ and $U(x)$ in such that the scaling (\ref{scale2}) is satisfied and the finite limits exist when $x\to+\infty$. In particular, if $c(x)\equiv 1$, than we take
 \be
 U(x) = 1+D\tanh(x/D), \quad x_2 = 0.
 \label{M4}
 \ee
 The results of calculations for $D=1$ ($U_{max}=2$) and $D=3$ ($U_{max}=4$)
 are shown in Fig.~\ref{f11}.

\begin{figure}[h!]
\vspace*{-5.cm}%
\centerline{\includegraphics[width=18cm]{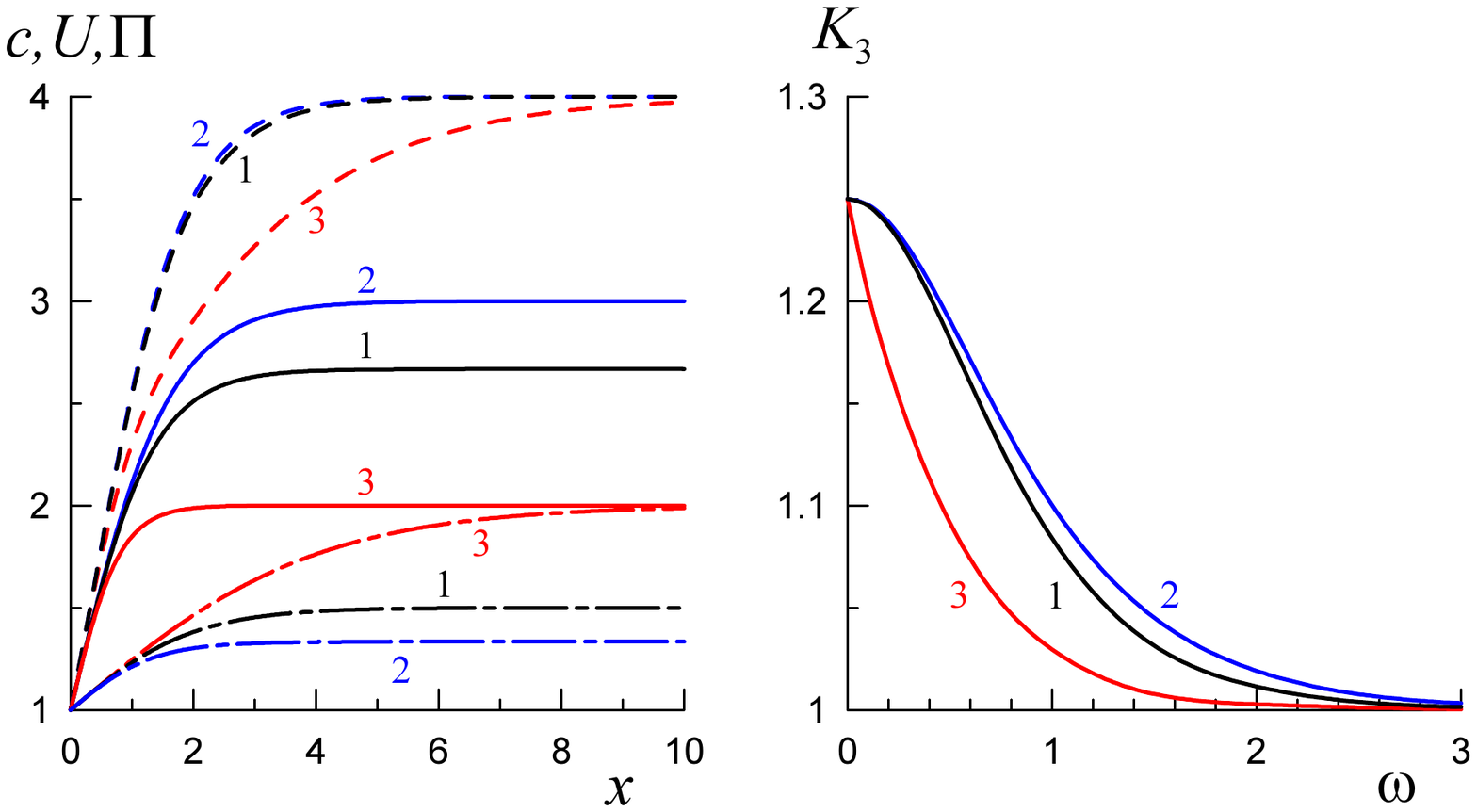}} %
\vspace*{-12.5cm}%
\caption{\protect\footnotesize (Color online) Left panel: $U(x)$ (solid lines),
$\Pi(x)$ (dashes), and $c(x)$ (dash-dotted lines) graphs.\ \ \ Right panel:
frequency dependence of the transmission ratio $K_3$.\ \ Figures indicate the
number of flow, see Eq.~(\ref{param}).}
\label{f12}
\end{figure}

 In the more interesting case when $c(x)$ and $U(x)$ grow simultaneously with $x$ (and $U(x) \ge c(x)$ everywhere), the calculations were carried out for current and wave speed in the following forms, respectively:
 \be
 U(x) = 1+A_U\tanh\left(\dfrac{x}{D_U}\right), \quad
 c(x) = 1+A_c\tanh\left(\dfrac{x}{D_c}\right),
 \label{M5}
 \ee
 with the various combinations of parameters satisfying the condition:
 \[
 \dfrac{A_U}{D_U} - \dfrac{A_c}{D_c}=1,
 \]
 which follows from the scaling (\ref{scale2}). Figure \ref{f12} presents the results of calculations for the three cases:
 \be
 \ba{lll}
 {\rm Case\ 1:}\ & \displaystyle A_U = \frac{5}{3},\ D_U = \frac{4}{3}, &  A_c = 0.5,\ D_c = 2, \\
 {\rm Case\ 2:}\ &  A_U = 2,\ D_U = 1.6, & \displaystyle A_c = \frac{1}{3},\ \ \ D_c = \frac{4}{3}, \\
 {\rm Case\ 3:}\ & \displaystyle  A_U = 1,\ D_U = 0.8, & A_c = 1,\ \ \ D_c = 4.
 \ea
 \label{param}
 \ee

 As one can see from Fig. \ref{f12}, the dependence $K_3(\om)$ in the Case~3 notably deviates from the dependences $K_3(\om)$ in the Cases 1 and 2. There are two distinction peculiarities in the Case~3. Firstly, the characteristic width of variation of the wave speed, $D_c$, significantly exceeds (5 times) that of current 
 variation $D_U$. Secondly, $U(+\infty) = c(+\infty)$, that is there is a third critical point at the infinity. The former distinction feature seems, however, not very important since even if $c(x)\equiv 1$,\ \ $K_3$ depends on $\om$ in the nearly the same way as in the Cases 1 and 2 (cf. line 2 and dashed line in the right panel of Fig.~\Ref{f11}). On the contrary, the presence of an additional (albeit infinitely remote) critical point has a profound impact on the behavior of function $r(x)$ which determines the progress of the wave transformation.
 
\begin{figure}[h!]
\vspace*{-2.5cm}%
\centerline{\includegraphics[width=18cm]{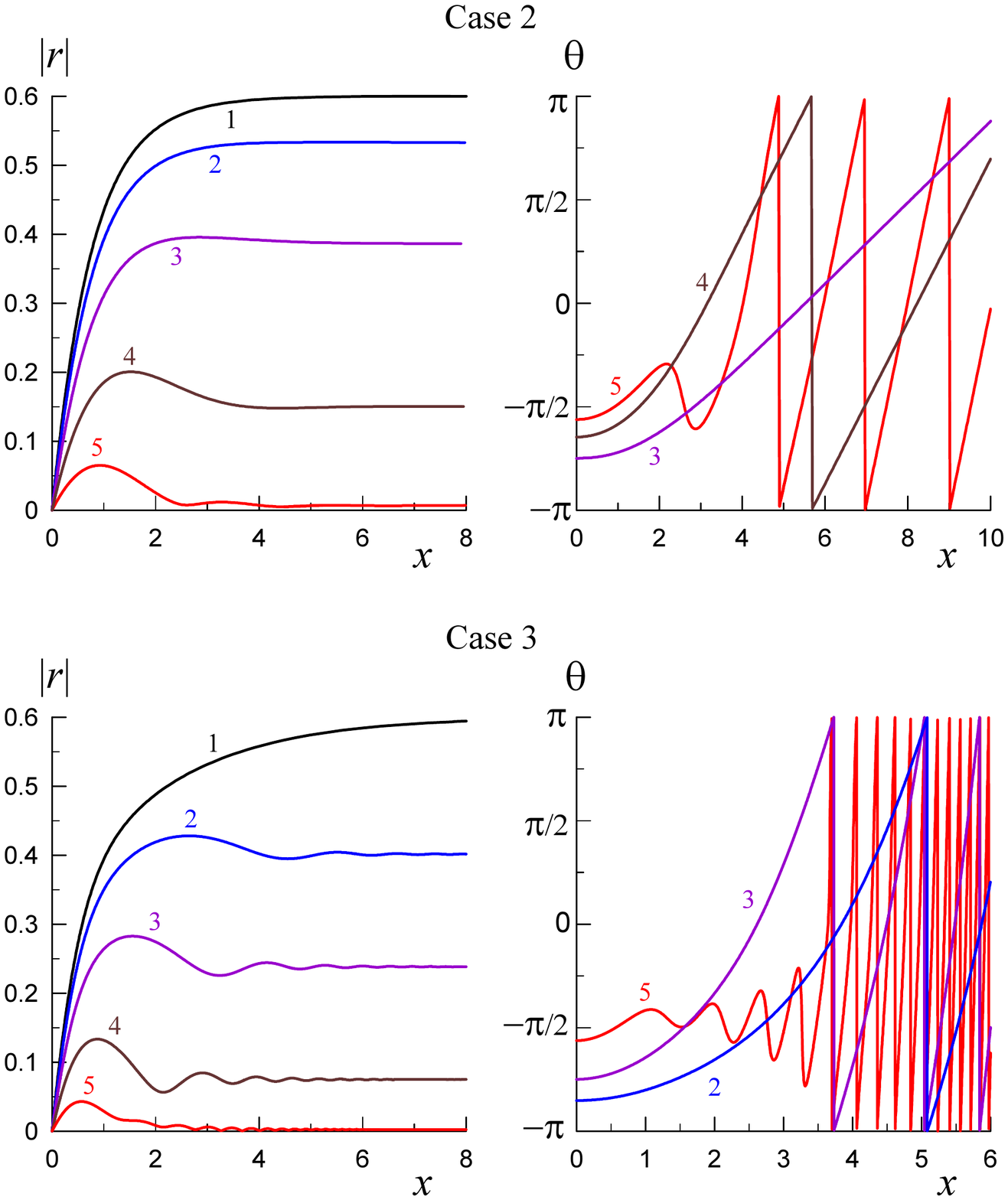}} %
\vspace*{-4.5cm}%
\caption{\protect\footnotesize (Color online) The absolute value and argument of the function
$r(x)$ for the Case~2 (top) and Case~3 (bottom). Different numbered curves pertain to different frequencies: \ \ 1 -- $\om=0$,\ \ 2 -- $\om=0.5$,\ \ 3 -- $\om=1$,\ \ 4 -- $\om=2$,
\ \ 5 -- $\om=5$.}
\label{f13}
\end{figure}

 Figure~\ref{f13} demonstrates the absolute value and argument of $r(x)$ for the Case~2 and Case~3; it allows us to compare the wave propagation in these two cases with the greater detail. It is clearly seen that in the Case~3 the oscillations of $r(x)$ begin to develop noticeably at a much smaller distance from the critical point
 $x_2 = 0$, and this results in lesser values and more pronounced oscillatory character of $|r(x)|$. Moreover, when $x$ increases, the oscillation period in the Case~3 decreases whereas in the Case~2 it approaches a constant value.

\section{DISCUSSION AND CONCLUSION}
\label{sec:7}
%
We have carried out the analysis of a simple
harmonic wave propagation in two types of ducts,
BH--WH and WH--BH, and have illustrated the theoretical results 
by numerical calculations. Note that in contrast to Ref. \cite{Peloquin} where the transformation of {\it dispersive} gravity-capillary waves were studied, in our paper devoted to {\it dispersionless} purely gravity long waves, each of the ducts is traversable only in the co-current direction, from left to right in the considered geometry. However, the problem statements for the aforementioned types of ducts are different and should be specified.

At the ends of a BH--WH duct far from the middle supercritical domain, the flow is subcritical (see Fig. \ref{f00}), and the problem statement is rather traditional. A wave running from the left in the course of propagation to the critical point $x_1$ is partially reflected due to scattering on the flow inhomogeneity.
A reflected wave runs to the left and, therefore does not have any influence on the wavefield downstream of the flow.
Upon attaining the point $x_1$, the positive energy incident wave passes through this critical point keeping its 
amplitude. In the supercritical domain at $x > x_1$ the wave propagates to the second critical point $x_2$ being amplified along the path due to the interaction with the negative energy wave (NEW) which is generated due to scattering of the positive energy wave on the flow inhomogeneity. In the result of such coupling, amplitudes of both waves of positive and negative energies grow simultaneously \cite{FabrStep}. The NEW  completely absorbs in the vicinity of the critical point $x_2$, whereas the wave of a positive energy passes through this point keeping its amplitude and penetrating into the right subcritical domain. In this domain, the penetrated wave propagates towards the infinity losing its energy for generation of reflected wave of positive energy due to flow inhomogeneity. Calculation of both the transmission ratio $K_2$ for the `active' duct domain with the supercritical flow and the transmission coefficient (ratio of the amplitudes of the transmitted and incident waves) do not cause any principal problem in this case. Note that the supercritical domain plays a role of an active zone of the open-boundary laser.

At the ends of a WH--BH duct far from the middle subcritical domain, the flow in contrast to the previous case is supercritical. Therefore, waves, both with positive and negative energies, entering the duct from the left end, propagate to the critical point $x_1$ interacting with each other in such a way that the total energy flux ${\cal E}$
determined by Eq.~(\ref{En2}) is conserved, whereas the amplitudes of both waves increase. In any case, only the finite domain to the left of the point $x_1$ is accessible for a observer, and in any point of it, one can see a superposition of both waves having no possibility to separate which of these two waves arrives from the left infinity. In such a situation, the only doubtless statement is, if only a positive energy wave is incoming from the left, then ${\cal E} > 0$, whereas the converse is not true, in general. However, as has been shown in the paper, in the neighbourhood of the critical point $ x_1 $, any NEW is completely absorbed. Therefore, it seems that the only reasonable statement of the problem in this case is to set the amplitude of the co-current propagating wave at either end of the subcritical domain and ignore the prehistory of wave propagation in the supercritical domain at $x < x_1$. Technically, it is more convenient to set the amplitude of the co-current propagating wave of positive energy at $x = x_2$, because both the reflected wave and NEW vanish in this point. When this is done, the picture of the wave propagation becomes quite clear. Integrating Eq. (\ref{r}) back from $x_2$ to $x_1$ and using Eq. (\ref{En2}), one can find the amplitudes of co-current propagating and reflected (counter-current propagating) waves at $x = x_1+0$ and determine what part of the wave energy penetrated from the left domain $x < x_1$ is lost in the middle domain between $x_1$ and $x_2$ due to the reflection on the flow inhomogeneity. Further, when the positive energy wave enters the supercritical right domain $x > x_2$, we can calculate its amplification due to the transformation into a NEW. This is precisely the statement of the problem that is
adopted in the paper.

Another interesting question is the choice of a flow
model to simulate a curved space-time, in particular,
the question what should be the effective metric at the
infinity. However, such problem statement is practically irrelevant for the WH--BH duct models because in this case the space
at the infinity is ``unusual'', with the metric signature
turned inside out. But for the BH--WH duct the 
choice of the model presumes, in general, that the
curvature vanishes at the infinity, and the analog
metric reduces to the Minkowski metric. At a first
glance, for
the metric as per Eq.~(\ref{G}) this requirement is equivalent to the condition $U(\pm \infty) = 0$, which makes the problem considered above incorrectly posed. Indeed, let $c(x)$ is finite everywhere and $U(x)\to 0$ when $x\to\pm\infty$. Then, first of all, we have a problem with the mass flux conservation Eq.~(\ref{Flux}). To cope with this problem, we need to assume that the duct width $W(x) \to \infty$ as $x \to \pm\infty$. Further, if a wave of a finite amplitude $A(-\infty) = A_0 \ne 0$ arrives from the left, then the perturbation (\ref{Sol}) remains finite at least up to the first critical point $x = x_1$. The energy flux (\ref{En2}) in this case will be finite (and constant) in this region too, including the far zone, $x \to -\infty$, where function $\Pi(x) = c(x)U(x) \to 0$. However, this is possible only in the case of the {\it total wave reflection}, $|r(-\infty)| = 1 $.
 It is interesting that the energy flux, in this case, can be quite different from zero and, being unchanged, is transferred to the next, supercritical, region. Then, after the inevitable transformation at the second critical point $x = x_2$, the energy flux can be transferred in the last, subcritical region. Here, in the case $U(+\infty) = 0$ the reflection coefficient must be also equal to one, which does not prevent the transmitted wave to have a nonzero amplitude at the infinity $A(+\infty)$. However, it is impossible, in this case, to find the transmission coefficient (the ratio of the amplitudes of transmitted and incidents waves) without additional assumptions.

 To avoid such a problem, we assume that the flow velocity $U(x)$
 is positive everywhere, and $c(x)$ and $U(x)$ have finite limits $c_\pm > U_\pm > 0$ when $x \to \pm \infty$, so that the analog space-time is flat at the infinity. The easiest way to verify this is to make a transformation to the new temporal coordinate (see, for example, \cite{Barc04}) $\dd \tau = \dd t - U \dd x /(c^2 - U^2)$, which reduces the interval to the Schwarzschild-like form:
 \[
 \dd s^2 = \left(\dfrac{G_2}{c\,U}\right)^2\left[-\Bl(c^2-U^2\Br)\dd\tau^2 + \dfrac{c^2\dd x^2}{c^2-U^2}+c^2\dd y^2\right].
 \]
Therefore, the flow models considered in this paper are quite suitable for the modelling of wave transformation in the curved space-time for both BH--WH and WH--BH duct models.

 The performed analysis has highlighted a very significant influence of the ``geometric factor'' $\Pi(x)$ on the wave propagation. Firstly, its constancy provides reflectionless wave propagation regardless of the frequency. Secondly, in the BH--WH ducts, the reflection coefficient in the subcritical flow regions depends mainly on the behavior of function $\Pi(x)$ rather than on the Mach number at the infinity, ${\cal M}_{\pm\infty} = U/c$ . In addition to that, amplification of positive energy waves in the supercritical region is also determined mainly by the difference in the values of $\Pi(x)$ at the ending points of the `active' (supercritical) region. With regard to this, we recall that at low frequencies, the gain does not depend which of the speeds is greater, $c_1$ or $c_2$ but is determined entirely by their ratio.
 
 In a WH--BH duct, the $\Pi(x)$ behavior has a profound impact on the wave propagation as well. However, the frequency dependence of the transmission ratio $K_3$ of its  final (supercritical) part is highly sensitive to that does the Mach number tend to unity at the infinity or not (see Fig.~\ref{f12}).

 In this paper we have considered quite general models of stationary flows variable in space. Such flows can be realised in laboratory water tanks with the varying tank width and depth. As has been shown, the metric provided by the flows for long surface waves is similar to the Schwarzschild-like metric in the general relativity. In the linear approximation, we have derived a second-order ODE describing the wave amplitude in the non-uniform flow with two critical points that mimic the BH and WH horizons in the general relativity. The derived ODE can be reduced to the set of two first-order ODEs containing critical points. Solutions to this set of equations can be investigated analytically in the vicinity of each critical point. This allowed us to determine the conditions of the optimal wave amplification for low- and high-frequency waves. The analytical results have been illustrated by the direct numerical solutions of the derived set of equation. 

In the next paper, we plan to present the detailed results on the wave amplification after passing through the `active' (supercritical) domain using exact analytical solutions on the wave propagation in the piece-wise linear background flow in shallow water. We will derive the transformation coefficients (the transmission and reflection coefficients) in the analytic form and quantify the results obtained in terms of the wave frequency, current speeds in the subcritical and supercritical domains, and length  of the supercritical domain. Ultimately  this should shed light on the intriguing problem of wave propagation through the wormholes if they exist in nature indeed. But, at least, the results obtained can be of interest to the interpretation of wave amplification in the natural estuaries (rivers and canals) or laboratory tanks. The validation of the theoretical results in the laboratory sets up is a challenge for the experimenters.

\begin{acknowledgments}
\noindent 
 S.C. was financially supported by the Ministry of
 Science and Higher Education of the Russian
 Federation.
 Y.S acknowledges the funding of this study provided by the grant No. FSWE-2020-0007 through the State task program in the sphere of scientific activity of the Ministry of Science and Higher Education of the Russian Federation, and the grant No. NSH-2685.2018.5 provided by the President of Russian Federation for the State support of leading Scientific Schools of the Russian Federation. The final version of this paper was completed when YS was a guest visitor of the Sydney University working on the SMRI grant in March--April 2021.
\end{acknowledgments}
 \appendix
 \section{Influence of the viscosity in the vicinity of critical points}
 To investigate the details of a solution in the vicinity of a critical point where $c(x) = U(x)$, let us take into account a small viscosity in the medium. Supplementing Eq. (\ref{Euler1}) with a viscous term and introducing the velocity potential $u=\partial\vp/\partial x$, we derive:
  \[
 \dfrac{\partial u}{\partial t}+\dfrac{\partial(Uu)}{\partial x}=-g\dfrac{\partial\eta}{\partial x}+
 \nu\,\dfrac{\partial^2u}{\partial x^2}\ \ \ \Longrightarrow\ \ \
 \dfrac{\partial\vp}{\partial t}+U\,\dfrac{\partial\vp}{\partial x}-
 \nu\,\dfrac{\partial^2\vp}{\partial x^2} = -g\eta,
 \]
 where $\nu\ll 1$ is the kinematic viscosity. Substitution of this into Eq. (\ref{eta}) yields:
 \[
 \ba{r}
 \nu\left(\dfrac{\partial}{\partial t}+U\,\dfrac{\partial}{\partial x}-2U\,\dfrac{c'}{c}\right)
 \dfrac{\partial^2\vp}{\partial x^2}+(c^2-U^2)\dfrac{\partial^2\vp}{\partial x^2}-
 2U\,\dfrac{\partial^2\vp}{\partial t\partial x}
   \\ \\
 +\,\left[2U^2\,\dfrac{c'}{c}- (c^2+U^2)\,\dfrac{U'}{U}\right]
 \dfrac{\partial\vp}{\partial x} - \dfrac{\partial^2\vp}{\partial t^2}+2U\,\dfrac{c'}{c}\,
 \dfrac{\partial\vp}{\partial t} = 0.
 \ea
 \]
 For a single harmonic, $\vp(x,t) = \phi(x)\re^{-\ri\,\om\, t}$, we have:
$$
 \nu U\,\dfrac{\dd^3\phi}{\dd x^3}+\left(c^2-U^2-2\nu U\,\dfrac{c'}{c} - \ri\,\om\,\nu\right)\dfrac{\dd^2\phi}{\dd x^2}
 $$
\be
 {} +\left[2U^2\,\dfrac{c'}{c} - (c^2+U^2)\,\dfrac{U'}{U} + 2\,\ri\,\om\, U\right]\dfrac{\dd\phi}{\dd x} + \left(\om^2 - 2\,\ri\,\om\, U \,\dfrac{c'}{c}\right)\phi = 0.
 \label{Visc}
 \ee
 
 In the neighborhood of the BH horizon (say, $x=x_1$), we put $x-x_1=\vep\xi$,
 where $\vep^2=\nu/(2\mu_1)$,\ \ $\mu_1= U'_1-c'_1>0$, and expand $c(x)$ and $U(x)$ in the Taylor series. Then Eq. (\ref{Visc}) yields in the leading order:
 \be
 \dfrac{\dd^3\phi}{\dd\xi^3}-\xi\,\dfrac{\dd^2\phi}{\dd\xi^2}-
 (1-\beta_1)\,\dfrac{\dd\phi}{\dd\xi}=0, \quad \beta_1=\frac{\om}{\mu_1}>0.
  \label{phi}
 \ee
 After integration of this equation over $\xi$, we obtain:
 \be
 \dfrac{\dd^2\phi_1}{\dd\xi^2}-\xi\,\dfrac{\dd\phi_1}{\dd\xi}
 +\ri\,\beta_1\,\phi_1=0, \qquad \phi_1=\phi-\phi_{01},\ \
 \phi_{01}={\rm const}.
 \label{phi1}
 \ee
 Then, by means of the ansatz $\phi_1(\xi) = \re^{\xi^2/4}G(\xi)$, we obtain the equation of the parabolic cylinder for the function $G(\xi)$:
 \be
 \frac{d^2G}{d\xi^2} + \left(\ri\,\beta_1 + \dfrac{1}{2} -
 \dfrac{\xi^2}{4}\right)G = 0.
 \label{G1}
 \ee
 Solution to this equation can be presented in terms of the functions ${\cal D}_{\ri\,\beta_1}(\pm\xi)$ and
 ${\cal D}_{-\ri\,\beta_1-1}(\pm\ri\,\xi)$ \cite{GR-07} so that, we obtain:
 \be
 \phi=\phi_{01}+\re^{\xi^2/4}\Bl[C_1{\cal D}_{\ri\,\beta_1}(\xi)+
 C_2{\cal D}_{\ri\,\beta_1}(-\xi)\Br].
  \label{ffi}
 \ee
 
This solution should be matched with the asymptotic expansions (\ref{asfi}) of the outer solution calculated for $x = x_1$. To this end, we use the asymptotics of the function ${\cal D}_p(s)$ for $|s| \gg 1$ (see \cite{GR-07}, Eq. 9.246):
 \bea
 {\cal D}_p(s) &\sim &
 s^p\re^{-s^2/4}{}_2F_0\left(-\frac{p}{2}\,,\frac{1-p}{2}\,;-\frac{2}{s^2}
 \right), \quad |\arg{(s)}| < \frac{3\pi}{4},
 \label{AssParCyl1} \\
 {\cal D}_p(s) &\sim & s^p\re^{-s^2/4}{}_2F_0\left(-\frac{p}{2}\,,
 \frac{1-p}{2}\,;-\frac{2}{s^2}\right)- \frac{\sqrt{2\pi}\,\re^{\ri\pi p}}
 {\Gamma(-p)}\,s^{-p-1}\re^{s^2/4}
 {}_2F_0\left(\frac{p}{2}\,,\frac{1+p}{2}\,;\frac{2}{s^2}\right),
 \label{AssParCyl2} \\
 {\cal D}_p(s) &\sim & s^p\re^{-s^2/4}{}_2F_0\left(-\frac{p}{2}\,,
 \frac{1-p}{2}\,;-\frac{2}{s^2}\right)- \frac{\sqrt{2\pi}\re^{-\ri\pi p}}
 {\Gamma(-p)}s^{-p-1}\re^{s^2/4}{}_2F_0
 \left(\frac{p}{2}\,,\frac{1+p}{2}\,;\frac{2}{s^2}\right), \phantom{WW}
 \label{AssParCyl3}
 \end{eqnarray}
 where Eq. (\ref{AssParCyl2}) is valid for $\pi/4<\arg{(s)} < 5\pi/4$, Eq. (\ref{AssParCyl3}) is valid for $-5\pi/4 < \arg{(s)} < -\pi/4$, $\Gamma(x)$ is the gamma-function, and
 \[
 {}_2F_0(a,b;z) \sim \sum_{n=0}^{\infty}\dfrac{\Gamma(a+n)\Gamma(b+n)z^n} {\Gamma(a)\Gamma(b)\,n!}.
 \]
 
 It can be easily seen that both terms in square brackets in Eq. (\ref{ffi}) grow infinitely (the former grows when $\xi\to-\infty$, and the later -- when $\xi\to+\infty$), therefore the matching with the bounded outer solutions (see Eq.~(\ref{asfi})) requires:
 \be
 C_1=C_2=0\ \ \Longrightarrow\ \ B^{(+)}_0= B^{(-)}_0=0, \quad
 A^{(+)}_0=A^{(-)}_0=\phi_{01}\re^{-\ri\,\psi(x_1)},
 \label{Match1}
 \ee
 because in the matching point $c_1 = 1$ (cf. Eq.~(\ref{sub})).
 
 In the neighborhood of the WH horizon (say, $x = x_2$), we put $x-x_2 = \vep\xi$, where $\vep^2=-\nu/(2\mu_2)$ because $\mu_2=U'_2-c'_2<0$. Then, we obtain the equation similar to Eq.~(\ref{phi1}):
 \[
 \dfrac{\dd^2\phi_2}{\dd\xi^2}+\xi\,\dfrac{\dd\phi_2}{\dd\xi}
 -\ri\beta_2\,\phi_2=0, \qquad \phi_2=\phi-\phi_{02},\ \
 \phi_{02}={\rm const},
 \]
 where $\beta_2=\om/\mu_2$. The ansatz $\phi_2(\xi) = \re^{-\xi^2/4}G(\xi)$ leads to the equation (cf. Eq. (\ref{G1})):
 \[
 \frac{d^2G}{d\xi^2} - \left(\ri\beta_2 + \dfrac{1}{2} +
 \dfrac{\xi^2}{4}\right)G = 0.
 \]
 Then, function $\phi$ can be written as:
 \[
 \phi = \phi_{02} + \re^{-\xi^2/4}\Bl[C_1{\cal D}_{-\ri\,\beta_2-1}(\xi) + C_2{\cal D}_{-\ri\,\beta_2-1}(-\xi)\Br].
 \]
This solution remains bounded for any finite constants $C_1$ and $C_2$. Indeed, assuming that $-\pi\le\arg{(\xi)} \le 0$ (and, consequently, $-\xi = \xi\re^{\ri\,\pi}$), using then Eqs.~ (\ref{AssParCyl1})--(\ref{AssParCyl3}), and keeping only the leading terms in each sum, we obtain:
 \[
 \phi\sim\phi_{02}+\left\{
 \ba{lr}
 \Bl(C_1 - C_2\re^{\pi\,\beta_2}\Br)\xi^{-\ri\,\beta_2-1}\re^{-\xi^2/2} + \dfrac{\sqrt{2\pi}\,C_2}{\Gamma(1 + \ri\,\beta_2)}\,\xi^{\ri\, \beta_2}, &
 \xi\to +\infty,
    \\ \\
 \Bl(C_2 - C_1\re^{-\pi\,\beta_2}\Br)|\xi|^{-\ri\,\beta_2-1}\re^{-\xi^2/2} + \dfrac{\sqrt{2\pi}\,C_1}{\Gamma(1 + \ri\,\beta_2)}\,|\xi|^{\ri\,\beta_2}, &
 \xi\to -\infty.
 \ea
 \right.
 \]
Match of this solution with the outer solution (\ref{asfi}) yields:
 \[
 A^{(+)}_0=A^{(-)}_0=\frac{\phi_{02}}{c_2}\,\re^{-\ri\,\psi(x_2)}
 \]
 and the relations between $C_1$ and $B^{(-)}_0$ on the one hand and $C_2$ and $B^{(+)}_0$ on the other demonstrating the fact that $B^{(\pm)}_0$ are mutually independent.

\end{document}